\documentclass[reprint,tikz,prb,twocolumn,superscriptaddress,showpacs]{revtex4-1} 
\AtBeginDocument{
\heavyrulewidth=.08em
\lightrulewidth=.05em
\cmidrulewidth=.03em
\belowrulesep=.65ex
\belowbottomsep=0pt
\aboverulesep=.4ex
\abovetopsep=0pt
\cmidrulesep=\doublerulesep
\cmidrulekern=.5em
\defaultaddspace=.5em
}

\usepackage{amsmath,amssymb,graphicx}
\usepackage{physics}
\usepackage{float}
\usepackage[usenames,dvipsnames]{xcolor}
\usepackage{tikz,amsthm,comment}
\usepackage{booktabs}
\usepackage[export]{adjustbox}
\usepackage[section]{placeins}
\usepackage[percent]{overpic}
\usepackage{microtype}
\usepackage{array}

\bibpunct{[}{]}{,}{n}{}{}
\usepackage[hidelinks]{hyperref} 
\hypersetup{colorlinks=true,citecolor=Red,linkcolor=Blue,urlcolor=Blue}

\definecolor{ForestGreen}{HTML}{668000}
\definecolor{red1}{HTML}{FF4136}
\definecolor{green1}{HTML}{00802b}

\newcommand{\<}{\langle}
\renewcommand{\>}{\rangle}

\begin{document}

\title{Gapless to gapless phase transitions in quantum spin chains}

\author{Shi Feng}
\affiliation{Department of Physics, The Ohio State University, Columbus, Ohio 43210, USA}
\author{Gonzalo Alvarez}
\affiliation{Computational Sciences and Engineering Division %
     and Center for Nanophase Materials Sciences, Oak Ridge National Laboratory, %
 \mbox{Oak Ridge, Tennessee 37831}, USA}
\author{Nandini Trivedi}
\affiliation{Department of Physics, The Ohio State University, Columbus, Ohio 43210, USA}

\begin{abstract}
We investigate spin chains with bilinear-biquadratic (BLBQ) spin interactions as a function of an applied magnetic field $h$. At the Uimin-Lai-Sutherland (ULS) critical point we find a gapless to gapless transition revealed by the dynamical structure factor $S(q,\omega)$ as a function of $h$. At $h=0$, the envelope of the lowest energy excitations goes soft at {\it two points} $q_1=2\pi/3$ and $q_2=4\pi/3$, dubbed the phase A. With increasing field, the spectral peaks at each of the gapless points bifurcate, making in total {\it four} soft modes, and combine to form a new set of excitations that soften at a {\it single} point $q=\pi$ at $h_{c1}\approx 0.94$. Beyond $h_{c1}$ the system enters another gapless B-phase until the transition at $h_{c2}=4$ to the fully polarized phase. We compare the ULS model results with those for the AKLT model as a representative of gapped Haldane phase.  We explain the mechanism of the gapless to gapless transition in the ULS model using its conserved charges and a spinon band picture. We also discuss the universality of central charges of the BLBQ family of models subjected to a magnetic field.
\end{abstract}
\date{\today}

\date{\today}
\maketitle

\section{Introduction} \label{sec:intro}
Quantum magnetism has been a subject of intense study, from exact solutions in one dimensions, to long range ordered states in higher dimensions, to quantum spin liquids arising from geometric frustration and competing interactions.   
Among various quantum magnetic systems, 1D spin systems are rather unique. In contrast to its higher dimensional counterparts, particles in one dimensional systems are highly affected by quantum fluctuations which prevent the breaking of continuous symmetries, and are much more likely to exhibit collective behavior because they cannot avoid the effects of interactions. 

One dimensional magnetic systems have a long history that dates back to 1931 when the exact solution of the spin-1/2 Heisenberg chain was found by Bethe \cite{Bethe1931}, predicting algebraic correlations in the ground state and gapless excitations. The mechanism of such gaplessness was given by the Lieb-Shultz-Mattis theorem whereby the separation between the ground and first excited state energies of a half-integer spin chain was shown to vanish in the thermodynamic limit \cite{LIEB1961407}. Haldane's generalization to larger spin-S SU(2) chains, using a mapping to a non-linear sigma model, showed that one dimensional Heisenberg antiferromagnets with integer spins have an excitation gap \cite{HALDANE1983464, PhysRevLett.50.1153, Affleck_1989}, later observed in experiments \cite{PhysRevLett.56.371,PhysRevB.38.543}. 
Following Haldane's prediction, much research has been done to study quantum phase transitions (QPTs) of integer spin chains under the influence of quadratic spin interactions and magnetic field \cite{Kiwata_1995,PhysRevB.58.R14709,PhysRevB.56.14435,PhysRevB.61.4019}. While these papers have provided some understanding of the magnetic properties of the BLBQ model, the static and dynamic properties of BLBQ models coupled to an external magnetic field have not been explored and is the topic of this paper.


The BLBQ Hamiltonian is a good description of (quasi) one-dimensional quantum magnetic systems such as CsNiCl$_3$ \cite{PhysRevLett.56.371,PhysRevB.42.4677,PhysRevLett.87.017202}, LiVGe$_2$O$_6$ \cite{PhysRevLett.83.4176,PhysRevLett.85.2380}. Recently we also proposed that such models naturally arise in strong spin-orbit coupled Mott insulators, such as OsCl$_4$, in which the transition metal is in the 5$d^4$ electronic configuration \cite{PhysRevB.91.054412,PhysRevB.101.155112}.

Our two main discoveries of the BLBQ spin-1 quantum chain as a function of $h$ are: (1) a continuous phase transition from the gapped Haldane phase to a gapless intermediate phase that precedes the polarized phase; and (2) a continuous phase transition from a gapless phase to another intermediate gapless phase for the Uimin-Lai-Sutherland (ULS)  critical Hamiltonian (see Eq.~\ref{eq:blbq}).
In case (2), while both phases harbor gapless excitations, their nature are different, with modes that go soft at different points in the Brillouin zone. There have been reports on electronic gapless to gapless phase transitions in metals that can be interpreted as a Lifshitz transition \cite{Lifshitz1960}, whereby the topology of the Fermi surface of the metal changes at the transition, resulting in a new metallic phase that gives rise to anomalies in the electronic properties~\cite{Rodney2013}.
We show that the QPT in the BLBQ model mentioned above can also be understood as a Lifshitz-type phase transition involving 3 spinon bands arising from the SU(3) symmetry at the ULS point, with 4 soft magnon modes decreasing to 1 soft mode across the transition. We compare the static and dynamical signatures of the field-induced phases of the BLBQ model at different points. We propose material candidates of BLBQ magnets where our predictions for the dynamical structure factor may be observable by inelastic neutron spectroscopy.
\begin{figure*}[tbh!]
    \centering
	\includegraphics[width=0.8\textwidth]{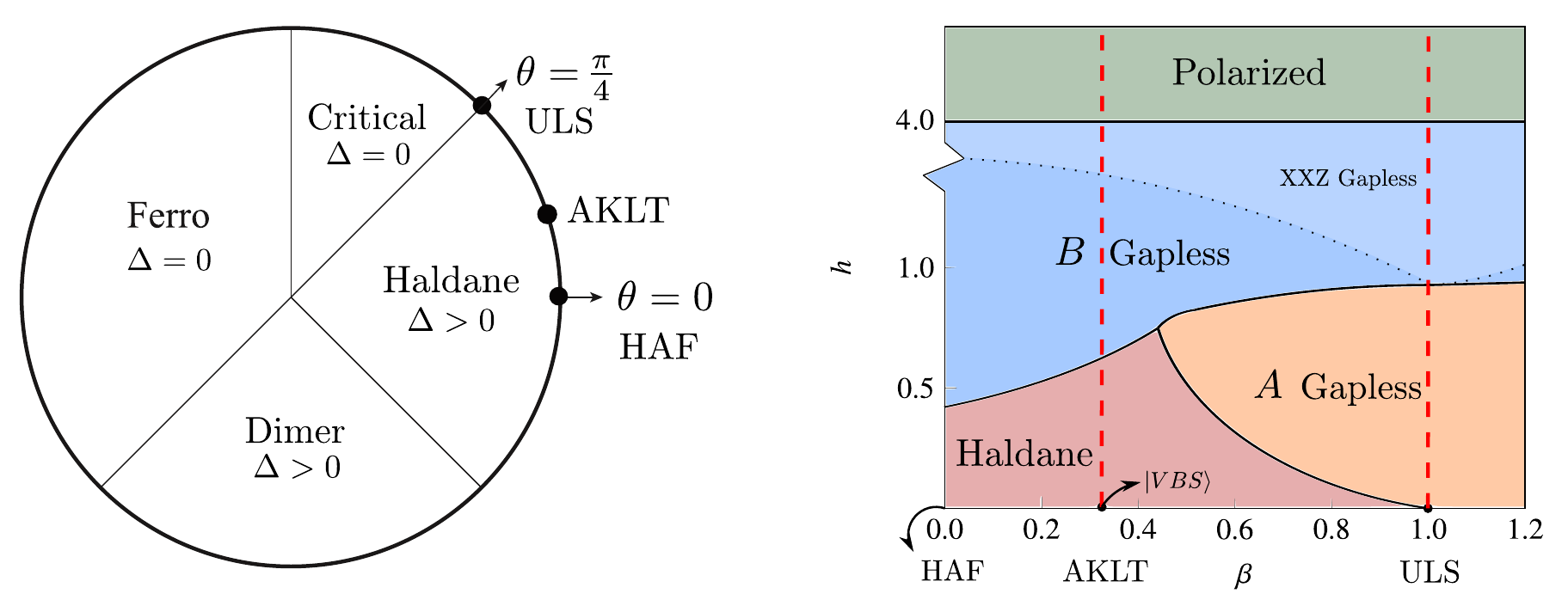}
    \caption{ \label{fig:phasedig} Left panel: phase diagram of the spin-1 bilinear-biquadratic (BLBQ) model parameterized by the angle $\theta$: $H = \sum_{\expval{ij}} \cos\theta\  \textbf{S}_i\cdot \textbf{S}_j + \sin\theta\ (\textbf{S}_i\cdot \textbf{S}_j)^2$. $\Delta$ denotes the gap which can be zero or finite in different phases. We focus on two representative  points: the Affleck-Kennedy-Lieb-Tasaki (AKLT) model at $\theta \approx 0.1024\pi$ ($\beta = 1/3$), and the critical ULS model at $\theta = \pi/4$ ($\beta = 1$). VBS refers to the valence bond solid ground state of the AKLT model at $h=0$. Right panel: schematic phase diagram of BLBQ parameterized by $\beta$ and $h$ reproduced from Ref.\cite{PhysRevB.58.R14709}. Phase boundaries are marked by black solid lines. black dashed line marks a cross-over to an effective spin-1/2 XXZ model in a field within the B  phase; at $\beta = 1$ the mapping is to an effective spin-1/2 Heisenberg model. We obtain the evolution of the static and dynamical correlation functions in the gapless phase A , B-phase  and Haldane phase along the two red dashed lines.}
\end{figure*}

The paper is organized as follows. Section II briefly reviews the BLBQ model and the phase diagram as a function of external magnetic field. Section III introduces definitions and computational methods. Section IV discusses the results for the AKLT model as a representative of the Haldane phase of the BLBQ, to be compared with those of the ULS model.
Our main results are shown in section V where we present both statics and dynamics of the ULS model and the phase transitions in a field. Section VI includes discussions of DMRG, the single mode approximation, extraction of the central charge of these models, and prospective materials to where our predictions may be observed. Section VI concludes with a summary and open questions.

\section{Model}
It was first argued by Haldane, and later rigorously proved,  that one dimensional Heisenberg antiferromagnets with integer spins have an excitation gap and finite correlation length \cite{HALDANE1983464,Affleck_1989}.
This gapped  one-dimensional integer-spin Heisenberg antiferromagnet  can be considered a particular case of the Haldane phase in a
more generic spin-1 bilinear biquadratic Hamiltonian (BLBQ) \cite{re:affleck86}, defined on a chain of $L$ sites by,
\begin{equation}\label{eq:blbq}
	H_{BLBQ} = \sum_{\expval{ij}}\textbf{S}_i \cdot \textbf{S}_j + \beta(\textbf{S}_i\cdot \textbf{S}_j)^2,
\end{equation}
where we have set the exchange energy $J = 1$. Its well-known phase diagram, parameterized by $\beta$ or the related angle $\tan\theta=\beta$, is shown in Fig.~\ref{fig:phasedig}.
In this paper, we will discuss the dynamical properties in these one-dimensional quantum magnets, particularly at the Affleck-Kennedy-Lieb-Tasaki (AKLT), Uimin-Lai-Sutherland (ULS), and Heisenberg points marked in the figure.
In addition, we add an external magnetic field $h$ yielding the Hamiltonian:
\begin{equation} \label{eq:hamiltonian1}
	H = H_{BLBQ} + h \sum_{i}S_i^z.
\end{equation}
where $h$ is measured in units of the exchange energy. 

We calculate the static and dynamical structure factors using the DMRG algorithm  \cite{re:white92, re:white93} for the Hamiltonian defined in Eq.~\ref{eq:hamiltonian1} to provide direct signatures that can be probed by neutron spectroscopy. Specifically we study $\beta = 1$ for the critical ULS model for which we find two transitions: a gapless to gapless transition at $h_{c1}$ and a second transition from a gapless to a polarized phase at $h_{c2}$. We contrast the behavior of the critical ULS point with the  AKLT model at $\beta = 1/3$ as a representative of Haldane phase that also shows two transitions but of different character: a gapped to gapless transition at $h_{c1}$, followed by a transition at $h_{c2}$ to a polarized phase.

Besides the unbiased DMRG results, we provide
interpretations of the gapless to gapless QPT using a spinon band picture. In the discussion section we apply a single mode approximation (SMA) analysis for the gapped-to-gapless transitions of AKLT Hamiltonian under a field, which shows the extent to which magnons in the Haldane and phase B can be captured by a single mode excitation, and indicate the degree of fractionalization. We also provide insights of universality of these phases via central charges. Finally we describe the candidate materials with 5$d^4$ electronic configuration and strong spin-orbit coupling that are suitable to observe the gapless-to-gapless phase transition in the orbital sector.

In the following, we define
$S_z \equiv  \sum_{i}S_i^z,$ and $E_\beta (S_z)$ the ground state energy of the BLBQ model at the parameter $\beta$ without a field in spin sector $S_z$.
Because both $H_{BLBQ}$ and the field term commute with $S_z$, $S_z$ is a conserved quantum number of $H$.
This implies that for every $h$, the ground state of Eq.~(\ref{eq:hamiltonian1}) with energy $E_\beta(h)$
is an eigenstate of  $H_{BLBQ}$ with energy $E_\beta(h)-h S_z$ for some $-L\le S_z\le L$. Moreover, this eigenstate
is the ground state of the sector or block of  $H_{BLBQ}$ with that value of $S_z$, so that by mapping
each $h$ to its $S_z$ sector, we can find the ground state of Eq.~(\ref{eq:hamiltonian1}) for any $h$.
Therefore, for a finite size system the QPT of the new Hamiltonian depends on the re-distribution of the energy spectrum of $H_{BLBQ}$: the QPT is driven by level crossings at certain $h_{c1}$ at which an old excited state becomes
the new ground state.  

To guide the discussion in this paper, we depict a schematic phase diagram of the BLBQ model at various values of $\beta$ for the BLBQ model in an applied magnetic field in Fig.~\ref{fig:phasedig} based on our DMRG results. We find that the gap in the Haldane phase closes at a critical field $h_{c1}$ and the system enters a gapless B-phase. In a small AKLT chain solved by ED (see Supplemental Fig.S7 \cite{re:supplemental}), this gaplessness can be viewed as a successive falling of excited states to the new ground states after the first level crossing at $h_{c1}$, and becomes a critical region in the thermodynamic limit. 

The system at the ULS point is gapless with 2 incommensurate soft modes, followed by the gapless phase A when subjected to a small magnetic field. We found that this phase A has 4 soft modes instead of 2 and persists for a large range of external fields before reaching the first critical point at $h_{c1}$.
At this point we find a gapless-to-gapless transition. 
In a small system solved by ED, the magnetization of the ULS model under a field exhibits steadily increasing steps, and is predicted to increase smoothly within the two phases in a large system as a function of $h$, until ultimately reaching
the transition point of the polarization field \cite{Parkinson_1989}. The calculation by DMRG shows more subtle structure prior to the gapless-to-gapless
phase transition at $h_{c1} \approx 0.94$, and that the magnetization is in fact zigzag instead of smooth even for large systems.
We will explain the behavior at this transition quantitatively by exploiting the SU(3) symmetry of the BLBQ model at the ULS point and developing a picture of Lifshitz-type transition that involves depopulation of spinon bands.


\section{Computational Methods} \label{sec:method}

\noindent \emph{Statics:} We first investigate the static signatures of the BLBQ model on a chain of $L$ sites
using density matrix renormalization group (DMRG) \cite{re:white92, re:white93}.
We calculate the spin-spin correlation function between spins separated by a distance $R$ defined by:

\begin{equation} \label{eq:RealSpCorr}
    \begin{split}
        C_S(R) &= \frac{1}{L} |\sum_{i} \langle  \mathbf{S}_{i} \cdot \mathbf{S}_{i+R} \rangle |,
    \end{split}
\end{equation}
%
where $i$ labels the sites. 
We also calculate the momentum-space correlations
%
\begin{equation} \label{eq:Ok}
    \begin{split}
        S(q) &= \frac{1}{L^2} \sum_{i,j} e^{iq(r_i - r_j)} \langle  \mathbf{S}_{i} \cdot \mathbf{S}_{j} \rangle ,
    \end{split}
\end{equation}
%
in order to elucidate the nature of the ground state. Here, $r_i$ and $r_j$ are the real-space
coordinates of sites $i$ and $j$, and $k$ represents the crystal momentum.
It is well-known that exponentially decaying spin-spin correlations indicate the presence of a spectral gap, whereas a power-law decay of correlations implies a gapless critical state~\cite{Hastings2006, Nachtergaele2006}.
Hence, although the static spin-spin correlations do not provide information about the dispersion of the modes,
they can nevertheless provide qualitative information about the nature of the ground state for varying external fields $h$.

\medskip

\noindent \emph{Dynamics:} 
The dynamical structure factor $S(q,\omega)$ as a function of 
frequency $\omega$ and momentum $q$ can be
measured with inelastic neutron scattering, adding to their importance. $S(q,\omega)$ is defined as usual
\begin{equation} \label{eq:sqw}
        S^{\alpha\beta}(q,\omega) = \frac{1}{L}\sum_{r} e^{-iqr} \int_{-\infty}^{\infty} dt \expval*{S_c^\alpha (t) S_{c+r}^{\beta}(0)} e^{i\omega t}
\end{equation}
which is related to Eq.(\ref{eq:Ok}) by $S(q) = \int S(q,\omega)d\omega$. 
To evalute Eq.(\ref{eq:sqw}) under open boundary condition (OBC) by DMRG, we take the central site $c$, and compute the dynamical structure factor by its analytic continuation which is given by the real space function:
\begin{equation}
S^{\alpha,\beta}(r, c, \omega) = \langle g.s.|S^\alpha_r \frac{1}{\omega + i\delta + H - E_0 }S^\beta_c|g.s.\rangle ,
\end{equation}
for all sites $r$, where $|g.s.\rangle $ is the ground state of the
Hamiltonian $H$ (either for the AKLT or ULS model), with or without magnetic
field, $E_0$ the corresponding ground state energy, and  $\delta$ a small broadening factor to ensure the convergence of the Green's function.
From the Fourier transform we obtain $S(q, \omega )$ and by integrating over all momenta, the density of states $S(\omega)\equiv S(c, c, \omega )$.

For $S_z=0$ or at $h=0$ the static and dynamic correlation functions involving $xx$, $yy$,
and $zz$ are all equal due to rotational symmetry. However, in a finite field, while $xx$ and $yy$ correlations remain equal, they can differ from the $zz$ correlations.
In what follows, we discuss the dynamical behavior of both $S^+S^-$ and $S^zS^z$ (the $S^zS^z$ dynamics are shown in the supplemental material \cite{re:supplemental}).

Reference \cite{re:arXiv160703538}
describes in detail our Krylov-space approach of dynamical DMRG.
The supplemental material \cite{re:supplemental} provides evidence of convergence with the number of states $m$ kept within DMRG, and shows when finite size effects in the dynamical structure factor can be neglected. 
We have used $\delta=0.05$ as the broadening factor, and have scanned the frequencies in increments of $\Delta \omega = 0.025$ in units of energy.
Both statics and dynamics are computed with DMRG with a desired truncation error $10^{-7}$ that requires us to retain up to a maximum number of $m = 800$ states.

\medskip

\noindent \emph{Entanglement:} 
The von Neumann entanglement entropy $S_{vN}$ also serves as another important signature of the model. $S_{vN}$ of a subsystem $A$ of the quantum spin chain with the rest of the chain is calculated by the reduced density matrix $\rho_A$:
\begin{align}
& \rho_A = \Tr_B\big[\ket{g.s.}\bra{g.s.}\big],\\
& S = -\Tr\big[ \rho_A \log (\rho_A)\big]
\end{align}
and provides a way to probe its entanglement structure. The second order transition point in a field is directly reflected in the discontinuity of entanglement entropy, which, in the low field regime, can be used as a benchmark especially for exactly solvable models like AKLT.

In addition, we also use entanglment properties to probe the possible conformal field theory (CFT) description of gapless modes. The entanglement entropy of 1+1 dimensional CFT under OBC satisfies
\begin{equation}
    S(n) = S^{CFT}(n) + S^{OSC}(n) + \textrm{const}
\end{equation}
where $n$ is the bond position. The first two terms $S^{CFT}(n)$ and $S^{OSC}(n)$ are defined as \cite{Calabrese_2004,PhysRevLett.104.095701} 
\begin{equation} \label{eq:cc}
\begin{split}
    	&S^{CFT}(n) = \frac{c}{6} \log\left[\frac{2L}{\pi }\sin(\frac{\pi n}{L})\right]\\
    	&S^{OSC}(n) = \sum_a F^a \left(\frac{n}{L}\right)\frac{\cos(2a\pi n /N)}{\abs{L\sin(n\pi/L)}^{\Delta_a}}
\end{split}
\end{equation}
where $c = N-1$ is the central charge and $\Delta_{a}$ the scaling dimension of the SU(N) Wess-Zumino-Witten (WZW) theory, $N$ defines the SU(N) symmetry of the effective CFT, $L$ the total length of chain and $F^a (n/L)$ is a universal scaling factor which has only one scaling dimension $a = 1$ for SU(2) and SU(3), and it can be treated approximately as a constant \cite{PhysRevB.92.054411,PhysRevB.94.195110}. We fit our data from DMRG against Eq.(\ref{eq:cc}) and extract the central charge $c$ in the gapless phases as an indicator of their universality class. Extracting central charge using Eq.(\ref{eq:cc}) involves the fitting of oscillatory waves with fine periodicity, therefore we have increased the number of states to $m=3000$ to enhance the accuracy of the fitting. With this value of $m$ we indeed obtain $c=2$ for the ULS model from Eq.(\ref{eq:cc}), exactly as expected by the SU(3) WZW theory.

\section{Haldane Phase}
\begin{figure*}[t]
{\centering \includegraphics[width=0.95\textwidth]{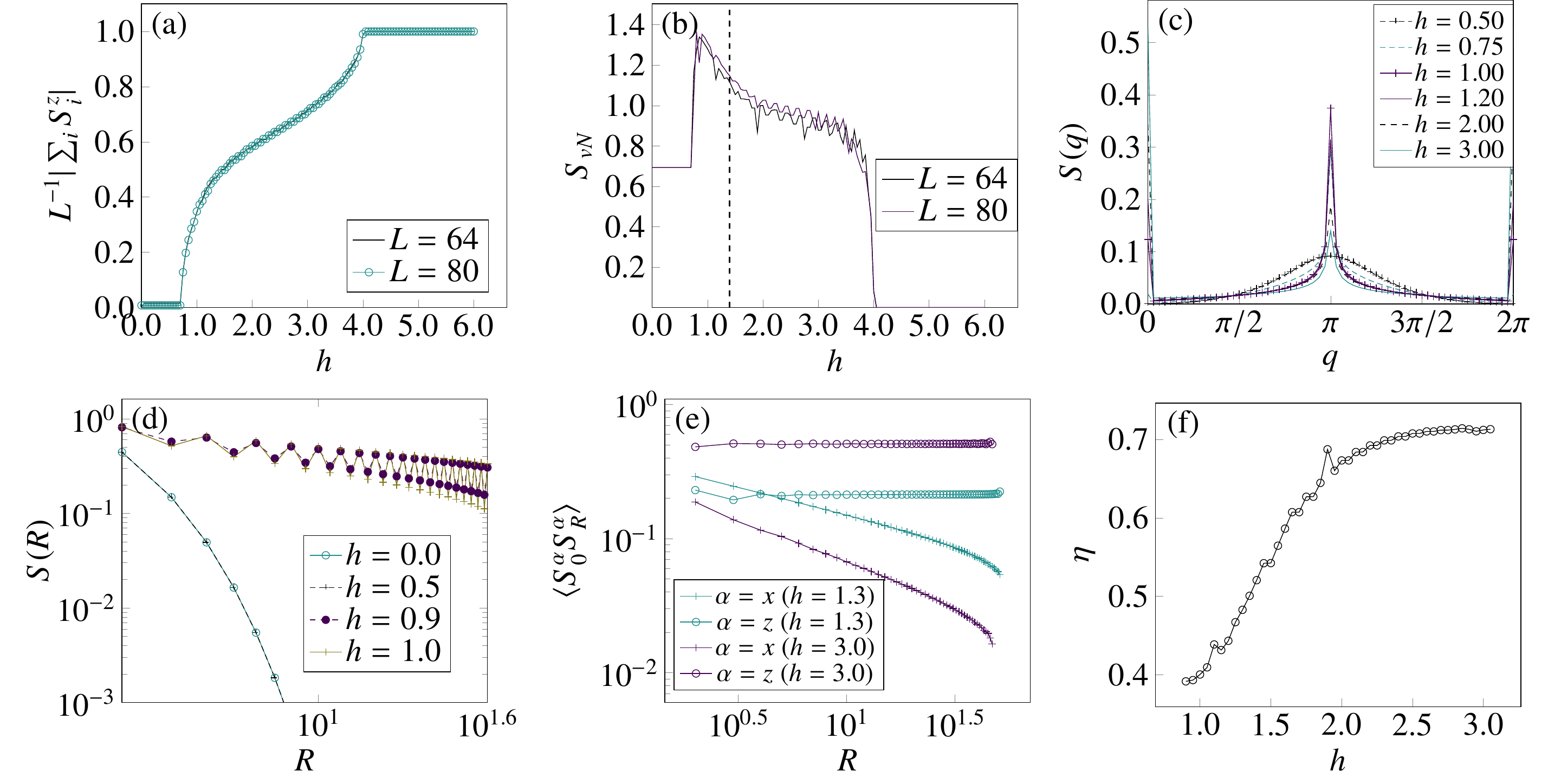}}
	\caption{\label{fig:akltall} Results of AKLT model under a field (a) Magnetization per site $s_z$ as a function of  $h \ge 0$, showing
    two second-order phase transitions. Data obtained from different system sizes converges rapidly and coincide with each other (b) von Neumann entanglement entropy $S_{vN}$ as a function of $h$ at central bond  computed for the same set of system sizes. 
    (c) Static structure factor under different external field. 
    (d) Real space correlation functions at $h = 0.0,0.5$ for the VBS state, and $h = 0.9, 1.0$ for the gapless phase B. Curves of the same phase coincide.
    (e) Correlation function of longitudinal and transverse components at different fields of phase B.
    (f) Exponent of real-space correlation function fitted by $S(R) \sim R^{-\eta}$ in the phase B.}
\end{figure*}

This section discusses the static and dynamical properties of the AKLT model, as a representative of the Haldane phase under an external field. It is defined by Eq.~(\ref{eq:blbq}) with
$\beta=\frac13$ and a ground state energy \cite{re:affleck88} $E_0/L = -2/3$. The Hamiltonian in an external field is:
\begin{equation}\label{eq:aklt}
	H_{AKLTZ} = \sum_{\expval{ij}}\textbf{S}_i \cdot \textbf{S}_j + \frac{1}{3}(\textbf{S}_i\cdot \textbf{S}_j)^2 + h \sum_i S_i^z,
\end{equation}
The AKLT Hamiltonian is not integrable. While some of its stationary eigenstates can be constructed explicitly \cite{AROVAS1989431,PhysRevLett.59.799}, less in known about the signatures of its excited states beyond the VBS ground state and about its dynamical properties, and are discussed below. 

\subsection{Statics of AKLT}  \label{sec:stataklt}
In this subsection we discuss the static behavior of an AKLT chain when subjected to magnetic field. We will look into its entanglement properties, magnetization and two-point correlations that probe phase transitions.  

The magnetization is obtained both by simulating the model with a field, and also by using the relation
\begin{equation}
h(S_z) = E(S_z+1) - E(S_z),\label{eq:hSz}
\end{equation}
with $E(S_z) = E_{aklt}(S_z)$ being the ground state energy of
the AKLT model in the $Sz$ symmetry sector \emph{without} the field. Since the total magnetization$S_z \equiv \sum_i S_i^z$ is a good quantum number, it can only increase in integer steps. As a result, we can compute quantum sectors of different $S_z$ separately, and the energy differences thereof can be attributed to different magnetic field $h(S_z)$. Figure~\ref{fig:akltall}(a) shows the magnetization per site vs magnetic field, where two critical points can be identified by kinks in the total magnetization.
Due to the non-zero gap above its $\ket{VBS}$ ground state, the zero-magnetization phase is protected before the gap is closed by the increasing magnetic field at $h_{c1}\approx 0.75\pm0.02$, after which the magnetization begins to increase until saturation at a polarization field $h_{c2} = 4$. 
Further evidence from
the von-Neumann entropy shown in Fig~\ref{fig:akltall}(b) also reflects the same transition.  

It is worth pointing out that $h_{c2} = 4$ marks the phase transition point to the polarized phase for all Hamiltonians in the BLBQ family. We briefly sketch the proof below: 
The critical value $h_{c_2}$ is the lowest field at which the BLBQ system becomes fully saturated.
For this to happen, the sector with  $S_z = L-1$ has to have lower energy than the sector with $S_z = L$, and the field needed satisfies
\begin{equation}\label{eq:lastSz}
E_{BLBQ}(L) - h_{c_2}L =E_{BLBQ}(L-1) - h_{c_2}(L-1),
\end{equation}
where $E_{BLBQ}(L')$ is the ground state energy of BLBQ Hamiltonian without field in sector  $S_z = L'$.
We use PBC, which coincides with OBC in the thermodynamic limit $L \rightarrow \infty$.
The sector with $S_z = L$ has only one state, with all spins having  $m = 1$ with the total energy contribution from  $H_{BLBQ}$ given by 
$E_{BLBQ}=(1+\beta)L$.
Now the sector with $S_z = L-1$ has exactly  $L$ states, and all states have  $L-1$ spins with  $m=1$ and one spin with  $m=0$. We call  $\ket{k_p}$ the state with  $m=0$ on the $p$-th site. Then
\begin{equation} \label{eq:hamkp}
H_{BLBQ} \ket{k_p} = \ket{k_{p+1}} + \ket{k_{p-1}} + [(1+\beta)L-2]\ket{k_p},
\end{equation}
which can be solved by a Fourier transform.
The $\ket{S_z = L-1}$ ground state can then be written as
\begin{equation} \label{eq:szlm1}
\ket{S_z = L-1} = \sum_{p} (-1)^p \ket{k_p},
\end{equation}
with energy $(1+\beta)L - 4$; using Eq.~(\ref{eq:lastSz}) yields $h_{c2} = 4$.
Moreover, the Von Neumann entropy of the ground state of the $S_z=L-1$ sector is exactly equal to $ln(2)$,
and that
of the fully saturated state is $0$. Therefore, the Von Neumann entropy has a discontinuity at $h=h_{c2}\equiv 4$, as expected,
due to the second order nature of the transition.
To conclude the phase diagram, we have identified three different phases: the SPT phase for $0 < h < h_{c1}$,
the gapless intermediate phase for $h_{c1} < h < 4$,
and the fully saturated phase for $h>4$, that we now further explore.

Note that in the thermodynamic limit $|h_{c1}| = \lim_{L\rightarrow\infty} |h_{c1}(L)|$, the results depend on the boundary conditions: for open boundary conditions (OBC),
$|h_{c1}(L)| = E_{aklt}(S_z=2) - E_{aklt}(S_z=1) > 0$, because the ground state
is four-fold degenerate with $Sz=0$ and $Sz=1$. For periodic boundary conditions (PBC),
the ground state is unique with $S_z=0$ and then
$|h_{c1}(L)| = E_{aklt}(S_z=1) - E_{aklt}(S_z=0) > 0$ \cite{re:affleck88}.

Figure~\ref{fig:akltall}(b) shows von Neumann entanglement entropy as a function of $h$ at the central bond of AKLT. At small fields the VBS ground state is unchanged, and $S_\rho = \log 2$ due to the pair of dangling spin-1/2 bonds at both ends. The Haldane-B transition  at $h_{c1}$ is evident by the sudden jump from the VBS plateau to a peak. Then $S_\rho$ drops at higher fields within the phase B and becomes zero beyond the transition at $h_{c2}$ to a product state. The decreasing $S_{vN}$ in phase B of AKLT is qualitatively different from that of the ULS model; in the latter it is a constant in the entire phase as shown in Sec.\ref{sec:uls}. Moreover, it is interesting that the $S_\rho$ of the AKLT model converges at high field close to $h_{c2} = 4$ to about the same value as $S_\rho$ of the ULS model. We can understand this behavior 
qualitatively in terms of a single mode approximation
as discussed in Sec.\ref{sec:discussion}.

The correlation functions of the AKLT model are shown in real space (Figure~\ref{fig:akltall}(d)) and in momentum space (Figure~\ref{fig:akltall}(c)) for different $h$.
In the VBS phase $ 0 < h < h_{c1} $, the ground state correlation function of Eq.(\ref{eq:aklt}) remains the same as that of the AKLT model's VBS state, because the field is not strong enough to change the nature of the ground state
from the $h=0$ VBS ground state.

\begin{figure*}[t]
{\centering \includegraphics[width=0.95\textwidth]{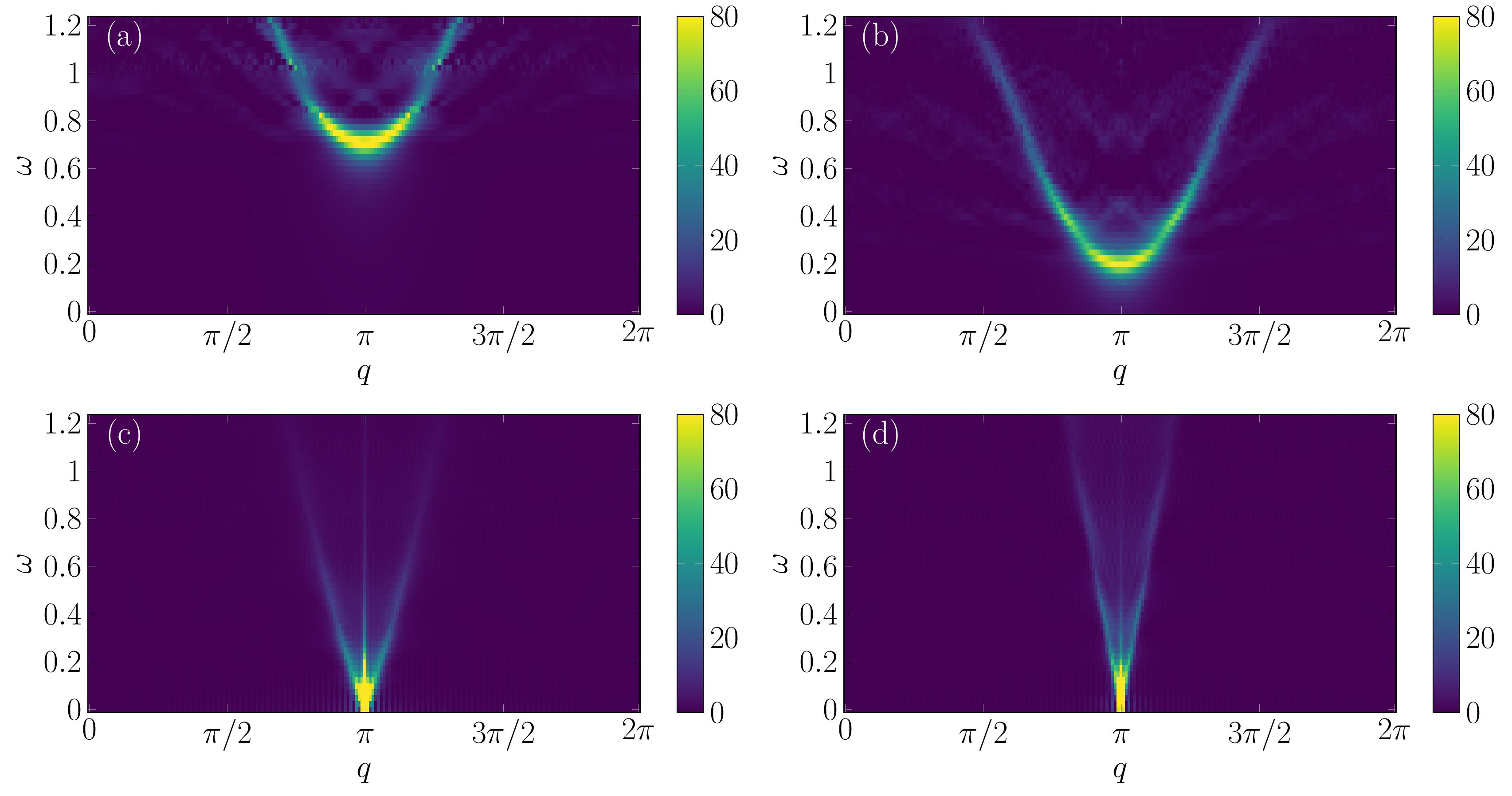}}
	\caption{\label{fig:akltdyn} Dynamic results of AKLT model under a field. $S^{+-}(q,\omega)$ at field $h=0$ ($sz=1/200$),
	$h=0.5$ ($sz=1/200$), $h=1.0$ ($sz\approx0.3$), and at field $h=1.5$ ($sz\approx0.5$). (a-c) are within phase A while (d) in within phase B. 
	Dynamical structure factors are obtained by 200-site DMRG under OBC.}
\end{figure*}
The two-point correlation function of a VBS state can be calculated analytically,
having an exponential behavior under OBC \cite{PhysRevLett.60.531}: 
\begin{equation} \label{eq:akltcorr}
\begin{split}
    &S^{\alpha\alpha}(r) = (-1)^n\expval{S_0^\alpha S_r^\alpha} = \frac{1}{3}\left(\frac{1}{3}\right)^{\abs{r}}\\
	&S^{\alpha\alpha}(q) = \frac{2(1-\cos(q))}{5+3\cos(q)}
\end{split}
\end{equation}
which is an exact result in the thermodynamic limit
arising from the hidden string order \cite{PhysRevB.40.4709,Girvin_1989}.
The finite correlation length and the static structure factor of the VBS phase in Eq.(\ref{eq:akltcorr}) is consistent with the fitting of our numerical data in Figure \ref{fig:akltall}(d):
the correlation function of the VBS ground state of spin-1 AKLT chain decays exponentially,
and the static structure factor has a smooth peak at $q = \pi$.
In the gapless intermediate phase, $h_{c1} < h < h_{c2}$, the correlations decay following a power-law correlation, indicating that the gap has vanished.
This is as expected, because the SPT phase cannot make a transition to a trivial phase without either breaking the symmetry or closing a gap.

The correlation function of the VBS Haldane phase is known from Eq.(\ref{eq:akltcorr}) analytically and agrees with our numerics. At higher fields in phase B, we have numerical results for the longitudinal and transverse correlations, Fig.\ref{fig:akltall}(d,e). 
By fitting the total $S(R)\propto R^{-\eta}$ or the transverse correlations for $h_{c1} < h < h_{c2}$ (since the longitudinal correlations are constant), we obtain the field dependence of the exponent $\eta$ shown in Fig.\ref{fig:akltall}(f).
We find that $\eta$ varies continuously with increasing field, which is reminiscent of the dependence of the exponent on the Tomonaga-Luttinger liquid interaction parameter. Also, as $h$ approaches $h_{c2}$,  $\eta\rightarrow 0.7$, which  is close to that of the ULS model at large field, supporting the claim of Fig.\ref{fig:phasedig} that both AKLT and ULS can be effectively captured by XXZ model when $h$ is close to (and smaller than) $h_{c2}$. 
This is discussed in greater detail in Sec.\ref{sec:uls}.


\subsection{Dynamics of AKLT}
We next present dynamical information of the AKLT model when subjected to a magnetic field, where we show explicitly the evolution of magnon bands with increasing field. Figure~\ref{fig:akltdyn} shows the dynamics in the $S^+S^-$ sectors, for different fields $h$, calculated using DMRG with the correction vector method \cite{re:kuhner99} on a $L=200$ OBC chain. (The $S^-S^+$ and $S^zS^z$ dynamics can be found in the supplemental material). These results should be compared with those of the ULS model to be discussed in the next section. 
For zero field, the $S^+S^-$ and $S^zS^z$ dynamics coincide, but start differing once the field is turned on since the field breaks time reversal symmetry.
For $h<h_{c1}$, the $S^zS^z$ dynamics is similar to that at $h=0$,
as can be confirmed analytically, because, (i) the ground state remains a VBS state, and,
(ii) $H-E_0$ does not depend on field, as the energy contribution of field in $H$ and $E_0$ cancels out.
On the other hand, the $S^+S^-$ dynamics
already shows a change: It moves down in energy exactly by $h$. This can also be confirmed analytically, because,
(i) the ground state remains the VBS state, and, (ii)
\begin{align}
\frac{1}{\omega - H(h) + E_0(h)}S^{\pm}_i|\textrm{VBS}\rangle =\nonumber\\
 \frac{1}{\bar{\omega}^\pm - H(h=0) + E_0(h=0)}S^{\pm}_i|\textrm{VBS}\rangle,
\end{align}
with $\bar{\omega}^\pm = \omega \pm h$, and
implies that the peak that is present  at $q=\pi$ for $h=0$ at $\omega = h_{c1}$
moves down (for $S^+$) \emph{linearly with $h$,}
so that at $h=h_{c1}$ it exactly touches $\omega = 0.$

For $h_{c1} < h < h_{c2}\equiv4$ (phase B), the $S^zS^z$ dynamics (see Fig.S3 of Supplemental \cite{re:supplemental}) has a peak at
$q=\pi$ and $\omega \approx h_{c1}$, a peak that decreases in intensity
as $h$ increases, and develops a FM peak that increases with increasing $h$ for $q = 0.$
Meanwhile,  $h_{c1} < h < h_{c2}\equiv4$, the $S^+S^-$ dynamics has a peak at $q=\pi$ and $\omega = 0,$
and two nearly linear branches of weak intensity, both going up in energy and away from $q=\pi$ to $q<\pi, \;q>\pi$:
one with negative slope and to $q>\pi$, and one with positive slope;
these branches slowly converge to each other and toward $q=\pi$ as the field $h$ goes to $h_{c1}$.
In other words, the slope of these branches slowly tends to infinity (become vertical) as $h$ increases to $h_{c1}$.
Moreover, as $h$ increases, the overall intensity of the $S^+S^-$  dynamics decreases, and becomes exactly zero at $h=h_{c2}\equiv4.$ It is worth pointing out that, the gapless mode at
$q = \pi$ in AKLT's phase B has a varying dispersion as magnetic field increases. This can be seen from Fig.\ref{fig:akltdyn}(c,d), where the dispersion is stretched to a wider energy range and the slope of the dispersion decreases, where the high energy tail gets heavier that reflects the increasing fractionalization from Fig.\ref{fig:akltdyn}(c) to (d). We explain this behavior in the discussoin below using the single mode approximation. It shows the dispersion at fields close to but smaller than $h_{c2}$ resembles that of the phase B of ULS ($\beta = 1$) and should be approximately the same as $q = \pi$ mode in Fig.\ref{fig:ulsdyn}(d). 

At even higher fields, we have the trivially ferromagnetic $h \ge h_{c2}\equiv4$ phase has no
$S^+S^-$ dynamics, but has non-zero $S^-S^+$ dynamics, and trivially FM
$S^zS^z$ dynamics  proportional to $\delta(\omega) \delta(k)$.


\section{ULS Critical Point} \label{sec:uls}
This section presents our main results on the QPTs in the Uimin-Lai-Sutherland (ULS)
model corresponds to the parameter $\beta=1$ of the BLBQ Hamiltonian family \cite{ULS1,ULS2,ULS3}. Under an external field the Hamiltonian is given by:
\begin{equation} \label{eq:ulsfermion1}
        H_{ULSZ} =  \sum_{\expval{ij}} \mathbf{S}_i \cdot \mathbf{S}_j + ( \mathbf{S}_i \cdot \mathbf{S}_j)^2 + h \sum_i S_i^z
\end{equation}
The ULS model has SU(3) symmetry, which is broken to U(1) $\times$ U(1) by the application of a magnetic field $h$ in the z-direction  \cite{PhysRevB.58.R14709,PhysRevB.101.155112}.
The ground state of the ULS model with an $h$ field then becomes the ground state of a block Hamiltonian with a fixed $S_z$ of the model without a field.

In the 70s, Uimin, Lai, and Sutherland used the Bethe ansatz method to describe the power law correlations in the ground state~\cite{ULS1,ULS2,ULS3}.
Kiwata~\cite{Kiwata_1995} studied the behavior of the ULS model under a magnetic field, and estimated the critical magnetic field $h_c$ at which the magnetization curve has a cusp, and showed that $h_c$ is a boundary between two states: the phase at lower fields containing excitations with $m = +1,0,-1$ and the higher field phase containing only $m=+1,0$. Later, F\'ath and Littlewood~\cite{PhysRevB.58.R14709} showed that in a field, one can identify a massless phase that is connected to the gapped Haldane phase and phase A with “depleting bands". These studies provide some intuition of distinct dynamics in each spinon sector. While the aforementioned works have provided a good understanding of the magnetic properties of the ULS Hamiltonian, their dynamics and critical behavior near the transition have not been explored. In this section we will discuss the relevant static phenomena first, followed by numerical and analytical analysis of its dynamics that lead to testable predictions for experiments. 

\subsection{Static Response of ULS Model}
\begin{figure*}[t]
{\centering \includegraphics[width=0.95\textwidth]{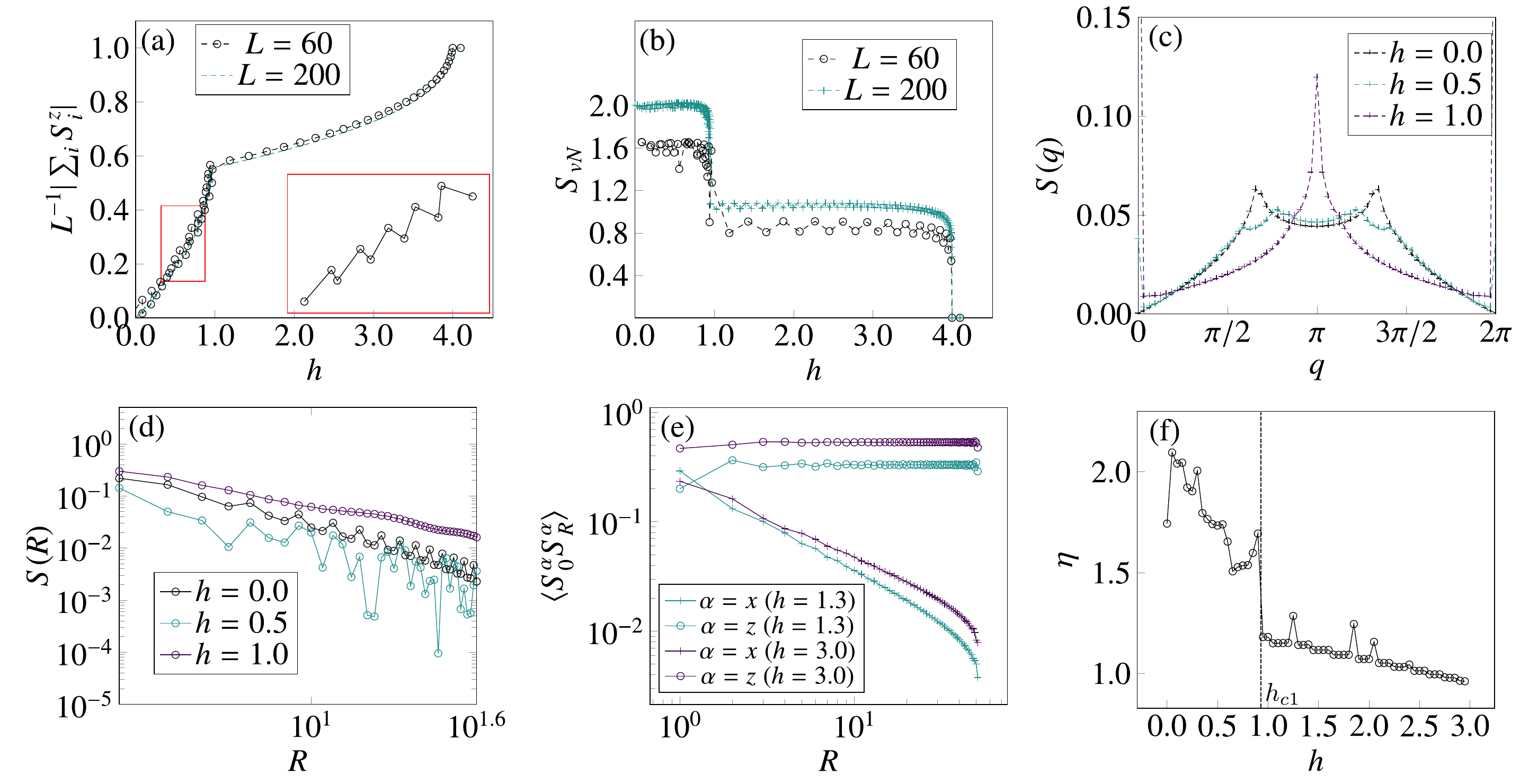}}
	\caption{\label{fig:ulsall}Results of ULS model under a field (a) Magnetization per site $s_z$ as a function of  $h$. Inset is the zoom-in segment of the magnetization in phase A which shows a zig-zag pattern regardless of system size.    
	(b) von-Neumann entanglement entropy $S_{vN}$ as a function of $h$ for a cut at the central bond for the same lattice and again computed with the DMRG. 
	(c) Static structure factor at $h = 0,0.5$ for the first phase, and $h = 1.0$ for the intermediate phase respectively
	(d) Real space correlation function for the same set of fields. 
    (e) Correlation function of longitudinal and transverse components at different fields of phase B.
	(f) Exponent of real-space correlation function fitted by $S(R) \sim R^{-\eta}$ for phase A and B separated by the vertical dashed line
	}
\end{figure*}
Figure \ref{fig:ulsall}(a) shows the magnetization as a function of $h$, where Eq.~(\ref{eq:ulsfermion1}) is evaluated for the ULS model using DMRG. Similar to the description in section \ref{sec:stataklt}, we use the relation $h(S^z) = E(S_z + 1) - E(S_z)$ to find the magnetization under different magnetic field $h$.
We see a second-order phase transition at $h_{c1}\approx0.94$.
As explained in previous section \ref{sec:stataklt}, the transition to the fully saturated phase occurs at $h_{c2}=4$.
Figure \ref{fig:ulsall}(b) shows the Von Neumann entanglement entropy obtained by integrating out half the system with a cut at the center bond, as a function of $h$.

The transition to the intermediate phase at $h_{c1}$ demands a different explanation from the one for the gapped AKLT model and other gapped models within the Haldane phase. In the ULS model there is no energy gap, hence it is not \emph{a priori} clear why the phase A of ULS is protected as $h$ increases. 
As we show next, the SU(3) symmetry of the ULS model can be exploited to explain the stability of the gapless phase A and the transition at $h_{c1}$.
For this purpose, it is helpful to map the ULS Hamiltonain onto a fermion model,
in which spin-1 operators are decomposed into partons by the mapping
$\mathbf{S}_i \equiv \psi_i^\dagger \mathbf{S}_i \psi_i$ with $\psi_i = (a_{i,1}, a_{i,0}, a_{i,-1})$
describing 3 annihilation components of a fermionic spinor corresponding to $m=1,0,-1$.
We follow the fermionizing approach of Ref. \cite{PhysRevLett.81.3527}
to show that the ULS Hamiltonian (with some auxiliary constants) can be written as
\begin{equation} \label{eq:ulsfermion2}
    \begin{split}
            H_{ULS} - const &=  \sum_{\expval{ij}} \mathbf{S}_i \cdot \mathbf{S}_j + ( \mathbf{S}_i \cdot \mathbf{S}_j)^2 - const\\
            &= - \sum_{\expval{ij}; mm'} a_{i,m}^\dagger a_{j,m} a_{j,m'}^\dagger a_{i,m'}
    \end{split}
\end{equation}
where we have defined the auxiliary constant term $const = n_i n_j + 3n_i$ with $n_i = \sum_{m}a^\dagger_{im} a_{im}$ being the total \emph{on-site} occupation number operator. This representation is faithful as long as there is 1 particle per site:
\begin{equation}
    \sum_{m=-1,0,1} a_{n,m}^\dagger a_{n,m} = 1
\end{equation}
(See appendix.\ref{sec:ulsfermion}). 
The fermionic representation helps understand if the system has a larger symmetry than apparent,
without having to write $H_{ULS}$ in terms of generators of the Lie algebra of SU(3).
Equation (\ref{eq:ulsfermion2}) can be compactly written as $-\sum_{\expval{ij}} (\psi_i^\dagger \psi_j) (\psi_j^\dagger \psi_i)$.
This  expression is invariant under any symmetry transformations in $\{ U \in GL(3,\mathbb{C})| U^\dagger U = \mathbb{I},\; \det(U) = 1\} = SU(3)$,
which directly shows the SU(3) symmetry of the ULS point.
From this representation, we see three conserved quantities, by computing the following commutators
\begin{equation}
    [\hat{N}_m, H_{ULS}] \equiv \left[\sum_i a_{i,m}^\dagger a_{i,m} , H_{ULS}\right] = 0,
\end{equation}
where we have defined $\hat{N}_m$ to be the total occupation number operator of $m$-type fermion of the whole lattice.
Let $N_m$ be the eigenvalue of $\hat{N}_m$, which must be an integer because it is a good quantum number.
We can hereafter identify
$N_{-1}$ as the number of sites with $m=-1$,
$N_0$ the number of sites with $m=0$, and
$N_1$ number of sites with $m=1$.
The ULS model then conserves $N_0$, $N_1$, and $N_{-1}$ \emph{separately}.
Because the sum of the three equals the number of sites $L$, there are \emph{two} linearly independent (l.i.) quantities;
thus the ULS model has \emph{two} local symmetries.

Let us choose the total $Sz$ and the total $N_1$: $[H_{ULS},Sz]$ = $[H_{ULS}, N_1] = 0.$
A field $h$ in the $z$ direction does not change these symmetries,
because $\sum_i S_i^z = \sum_i a_{i,1}^\dagger a_{i,1} -  a_{i,-1}^\dagger a_{i,-1}$ obviously commutes with all $N_m$.
We can then label the energy of the ULS in each block $(Sz, N_1)$ with
$E_{ULS}(Sz, N_1)$, and the energy of the ULS \emph{with} field as $E_{ULS}(Sz, N_1) + hSz$.
Because we choose $h \ge 0,$ $N_1$ tends to decrease as $h$ increases, and $Sz$ tends to become
more negative, so that $|Sz|$ increases as $h$ increases.
For $h < hc_1,$ the system can decrease its energy by either increasing $|Sz|$, or, by decreasing $N_1$
(because both are conserved and l.i.), or, both.
Decreasing $N_1$ while at the same time increasing $|Sz|$ increases $E_{ULS}(Sz, N1)$
but decreases $hSz$, so the two terms compete.

At first, it costs more to constantly increase $|Sz|$, and the system must instead zigzag $|Sz|$.
But at some large enough field $h$, the field term wins and decreasing
$|Sz|$ is no longer advantageous energetically.
This happens when $N_1$ cannot be decreased any further, that is, when $N_1$ reaches its minimum value: zero.
This point marks the second order phase transition at $h=h_{c_1}$.
From $h>h_{c1}$ onward, there are no longer ground states with $m=1$ sites,
 the magnetization $Sz$ equals $-N_{-1},$ and $N_1 = 0.$
 
Figure.\ref{fig:ulsall}(d,e) shows the numerical results of the decay of the real space correlations (both longitudinal and transverse) for the ULS model subjected to different fields. The power law exponent $\eta(h)$, as shown in Fig.\ref{fig:ulsall}(f), varies continuously in both phase A and phase Bs but changes dramatically at transition $h_{c1}$, indicating an abrupt change in the underlying Tomonaga-Luttinger theory. As the field increases toward $h_{c2}$, $\eta$ gradually decreases and converges $\eta \rightarrow 0.7$, consistent with the behavior in the AKLT model. 

Figure.\ref{fig:ulsall}(e) shows the power-law decay of longitudinal components $S_0^z S_R^z$ of phase B, which is again almost constant, thus the decay of $S(R)$ is mainly attributed to the transverse components $S_0^x S_R^x$ and $S_0^y S_R^y$ like in the AKLT model. This behavior can be quantitatively described by exploiting the SU(3) symmetry at ULS point and its spinon bands. In the fermion representation, $S_z = n_1 - n_{-1}$, hence the longitudinal correlator is $\expval{S_i^z S_j^z} = \expval{n_{i,1}n_{j,1}} + \expval{n_{i,-1}n_{j,-1}} - \expval{n_{i,1}n_{j,-1}} - \expval{n_{i,-1}n_{j,1}}$. Noting that spinon of 1-type is completely depleted in phase B, the longitudinal correlator is reduced to
\begin{equation}
	\expval{S_i^z S_j^z} = \expval{n_{i,-1} n_{j,-1}}
\end{equation}
for the ULS model in an intermediate field. Also Fig.\ref{fig:ulsall}(e) shows that, within phase B, the transverse correlator is almost constant, indicating that  $n_{-1}$ in the phase B is approximately ordered with
\begin{equation}
	\expval{n_{-1}} \simeq \sqrt{\expval*{S_i^z S_j^z}},\;\;\; \forall\; 0< i,j \le L
\end{equation}
Using $\sum_m\expval{n_{i,m}} = 1$ and the fact that $n_1 = 0$ in phase B, we must have
\begin{equation}
	\expval{n_{0}} \simeq 1 - \sqrt{\expval*{S_i^z S_j^z}}
\end{equation}
implying that $n_0$ is also ordered in phase B. In fact for the polarized phase $h>h_{c2}$, our numerical calculation indeed gives $\expval*{S_i^z S_j^z} = 1,\;\forall\; 0<i,j\le L$, consistent with $\expval{n_{-1}} = 1$ and complete depletion $\expval{n_0} = 0$. 
This provides a description of the phase transition via depopulation of bands and its resemblance to Lifshitz transition.  

Now that $\expval*{S_i^z S_j^z}$ is a constant in phase B (both $n_0$ and  $n_{-1}$ are ordered), the decay of $S(R) = \expval{\mathbf{S}_0\cdot\mathbf{S}_R}$ is entirely attributed to transverse spin components like $\expval*{S_i^x S_j^x}$, which arise from the exchange of particles between the two spinon bands. It is simple to check that within phase B the transverse contribution is related to a kinetic exchange of spinons among two flavors at a given site, described by,
\begin{equation}
    \begin{split}
        \expval{S_i^x S_j^x} &=\expval{K_i K_j} \\
        K_i \equiv \frac{1}{\sqrt{2}} \Big(a_{i,0}^\dagger &a_{i,-1}^{\phantom{\dagger}} + a_{i,0}^{\phantom{\dagger}}a_{i,-1}^\dagger \Big)
    \end{split}
\end{equation}
Hence even though spinons are ordered in the orginal lattice, transverse correlations of spins nonetheless show a power law decay.

\begin{figure*}[t]
{\centering \includegraphics[width=0.95\textwidth]{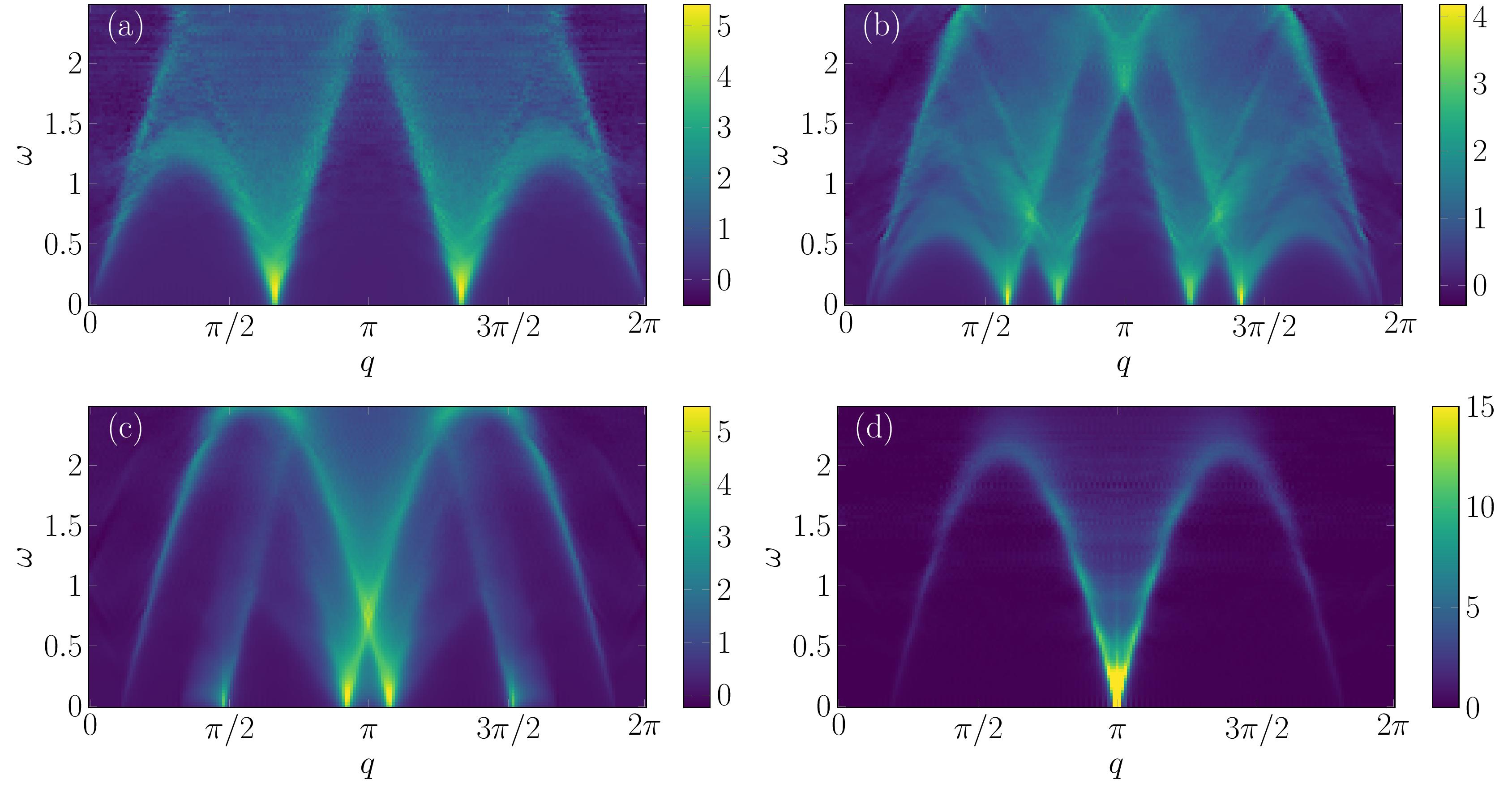}}
	\caption{\label{fig:ulsdyn} $S^{+ -}(q,\omega)$ dynamics of ULS model at field $h=0$ ($sz=1/200$)
    , $h=0.5$ ($s_z \approx 0.16$), $h=0.9$ ($s_z \approx 0.50$), and $h=1.5$ ($s_z\approx 0.60$).
    (a-c) are within phase A while (d) is within phase B. Results of dynamics are obtained by 200-site DMRG under under OBC.}
\end{figure*}

\subsection{Dynamics of ULS Model}

It was shown in Ref. \cite{PhysRevB.61.4019} that at $\beta = 1$, the spin-1 chain has an exact mapping to a Schwinger boson representation by projecting out the antiparallel states in the bond-operator representation at large enough magnetic field before saturation at $h_{c2}$. Thus in a large enough magnetic field the spin-1 chain can be considered a spin-1/2 chain. This boson representation gives a very good picture for understanding the magnetization of spin-1 ULS at large field qualitatively. There are two more questions we can ask based on this insight: how does the SU(3) system continuously transit to an effective SU(2) system, and, how can we describe the dynamical evolution from a spin-1 chain to its effective spin-1/2 map. In this subsection we will discuss these questions using the results of the dynamical correlations for the ULS model obtained from DMRG.

Figure \ref{fig:ulsdyn} shows the $S^+S^-$ component of the dynamical structure factor calculated using DMRG for a lattice of 200 sites with
open boundary conditions, with and without field a $h$, as indicated.
(The $S^zS^z$ and $S^-S^+$ components are shown in the  Supplement \cite{re:supplemental}).
Before the first transition at $h_{c1}$, the dynamical structure factor $S(q,\omega)$ in  Fig.~\ref{fig:ulsdyn}(a) shows that a wide range of frequencies are excited at a given momentum, 
in contrast to the gapped Haldane phase that shows sharper modes.  At low energies, the spectrum for the ULS model has two gapless incommensurate modes at $q \sim 2\pi/3, 4\pi/3$ corresponding to the two peaks shown in Fig.~\ref{fig:ulsall}(c) at $h = 0$. This is a distinct
fingerprint observable by inelastic neutron-scattering. The broad spectrum provides clear evidence for
fractionalized excitations.

Adding a field breaks the SU(3) symmetry of the ULS model into U(1)$\times$U(1). As shown in Fig.~\ref{fig:ulsdyn}(b), this reduction of symmetry is accompanied by the
bifurcation of the two incommensurate modes that are both two-fold degenerate, resulting in 4 distinct gapless modes. Upon increasing the field, we find that
the two pairs of modes move in opposite directions in momentum space. Near $h_{c1}$ one pair of modes recombines into a degenerate mode at
$\pi$, the other pair moves further away from each other and becomes fainter as the field reaches $h_{c1}$.  Finally, as shown in 
Fig.~\ref{fig:ulsdyn}(d), at $h  = 1.5 > h_{c1}$ we see only one gapless mode at $\pi$ while the other pair is completely washed out.     

Importantly, in Fig.~\ref{fig:ulsdyn}(h) which shows the dynamical structure factor at $h = 1.0$, the $S(q,\omega)$ is not as sharp and linear as in the AKLT model or Heisenberg model shown in the Supplement\cite{re:supplemental}, instead it forms a fan emanating from the gapless point at $q = \pi$ to higher energies with decreasing spectral weight. This behavior resembles the spectrum of a spin-1/2 Heisenberg chain obtained by Bethe ansatz \cite{Bethe1931,Mourigal2013,PhysRevLett.111.137205}.

In order to understand how the phase transition is reflected in the bifurcation in dynamical structure factor shown in Fig.\ref{fig:ulsdyn}(a-d), we adopt a fermion band representation of the problem.  
Let $k_m$ denote the fermi momentum of the spinon of $m$-flavor. The single-particle density of $m$-flavor spinon can thus be approximated by $\rho_m \simeq \int_{-k_m}^{k_m}dk/2\pi = k_m / \pi$. Notice that the spinon representation is faithful iff $n_i \equiv 1$, and that $\rho_m =  \sum_{i} \expval*{n_{i,m}}/N$, the 3 fermi momenta are thus related by
\begin{equation} \label{eq:constraint}
	\sum_{m=-1,0,1} k_m = \pi
\end{equation}
The magnetic field contribution in terms of spinons is
\begin{equation} \label{eq:hSz}
    hS_z = \sum_i h(a_{i,1}^\dagger a_{i,1}^{\phantom \dagger} - a_{i,-1}^\dagger a_{i,-1}^{\phantom \dagger})
\end{equation}
without which all 3 bands are degenerate, hence on the ULS point $k_m = \pi/3$ for all 3 bands. As SU(3) is broken by a small but non-zero field, $k_0$ will remain intact, yet the other two momenta will change by $k_{1,3} = \pi/3 \pm h/v$,
where $v$ is the spinon velocity.   

Next we show this fermion band picture provides an explanation of the dynamical spectral function shown in Fig.\ref{fig:ulsdyn}. Low energy spinons of the SU(3) model can be approximated by a pair of chiral fermions \cite{AFFLECK1988582}
\begin{equation}
	a_{i,m} \approx f_{L,m}(x) e^{-ik_m x} + f_{R,m}(x) e^{ik_m x}
\end{equation}
where $f_{L,m}$ and $f_{R,m}$ respectively denote left and right chiral fermion annihilation operators relevant for $m$-spinon with momenta $k_m = \pi/3$. Therefore, in the low energy sector for $h \ll vk_m$, the magnon excitation can be approximated by
\begin{equation}
	\begin{split}
		S^+(x) &\approx f_{R,1}^\dagger f_{L,0}^{\phantom \dagger} e^{-i(k_1 + k_0)x} + f_{L,1}^\dagger f_{R,0}^{\phantom \dagger} e^{i(k_1 + k_0)x}\\
		       &+ f_{R,0}^\dagger f_{L,-1}^{\phantom \dagger} e^{-i(k_0 + k_1)x} + f_{L,-1}^\dagger f_{R,0}^{\phantom \dagger}e^{i(k_1 + k_0)x}
	\end{split}
\end{equation}
in terms of the scattering channels between left and right chiral fermions. From previous analyses of fermion bands, it is readily seen that the momenta relevant for these processes are
\begin{equation}
\begin{split}
    	&k_1 + k_0 = 2\pi/3 - h/v,\\
	&k_0 + k_{-1} = 2\pi /3 + h/v\  \cdot
\end{split}
\end{equation}
This explains the bifurcation of modes at $q = 2\pi/3$ and $q = 4\pi/3$ shown in Fig.\ref{fig:ulsdyn}(f). Further increasing $h$ towards $h_{c1}$ leads to the reduction of fermi momentum $k_1$, thus the de-population of spinon of $m=1$ type. Its complete de-population happens at $h = h_{c1}$ - exactly the end of the zigzag magnetization pattern. In other words, $h_{c1}$ can be viewed as the chemical potential $\mu_1 \equiv h_{c1}$ of the 1-type spinon, which touches the bottom of the 1-type spinon band and gives a zero occupation at ground state. Therefore upon entering the B-phase, all excitation channels in $S^+$ relevant for $f_{R/L,1}$ vanish, and the only modes left are those with $k_{1} + k_0 = k_0 = \pi$. This explains the “recombination" of modes shown in Fig.\ref{fig:ulsdyn}(a-d). 

Furthermore, this fermionic band picture also allows us to explain the square-root-like scaling behavior in the magnetization for $0 < h_{c1} - h = \delta h < h_{c1}$ near the critical point $h_{c1}$. As is clear from Eq.(\ref{eq:hSz}), varying $h$ is equivalent to a varying chemical potential $\mu_1'(h)$ of relevant spinon. Its dispersion for small $\delta h > 0$ can then be written as
\begin{equation}
    \epsilon_{1,q} - \mu_1' = \epsilon_{1,q} - \delta h \  \cdot
\end{equation}
Assuming a parabolic dispersion $\epsilon_{1,q} = \alpha q^2$ of the 1-type spinon near the bottom of the band, where $\alpha$ is a constant, at the Fermi momentum we have
\begin{equation}
    k_1 = (\delta h / \alpha)^{1/2}
\end{equation}
From Eq.(\ref{eq:hSz}) it's magnetization near $h_{c1}$ can be evaluated by
\begin{equation}
    S_z = \abs{k_1 - k_{-1}} = k_{-1} - (\delta h / \alpha)^{1/2}
\end{equation}
which immediately determines the critical exponent of magnetization near $h_{c1}$:
\begin{equation} \label{eq:criticalhc1}
    \delta S_z(h_{c1}) \propto \delta h^{1/2}
\end{equation}
which agrees with numerical results in Fig.\ref{fig:ulsall}(a). The same physics takes place at the second phase transition near $h_{c2} = 4$, where, instead of 1-type spinon, it is the 0-type spinon that gets depopulated due to the shift of its fermi momentum $k_0$, or equivalently its chemical potential $\mu_0$. Because of the complete depopulation of 1-type spinon at $h_{c1}$, the constraint of Eq.(\ref{eq:constraint}) changes into $k_0 + k_{-1} = \pi$. The Zeeman term relevant for (-1)-type spinon $-h a_{n,-1}^\dagger a_{n,-1}$ raises the chemical potential $\mu_0$, thus continuously lowers the energy of the lowest occupied state. This leads to the transfer of spinons from $m=0$ into $m=-1$ band. Assuming a parabolic band of 0-type spinon again gives the same magnetic critical exponent $S_z(h_{c2}) \propto \delta h^{1/2}$. Ultimately at $h_{c2}$ the $m=0$ band becomes completely unoccupied and we obtain $k_{-1} = \pi$ as in Fig.\ref{fig:ulsdyn}(d). 

Moreover, the magnetic critical exponent on the right side of $h_{c1}$ ($h_{c2} > h > h_{c1}$) is readily derived starting from the phase B. Noting that $h\sum_i S_i^z$ is a good quantum number, previous analysis on $S_z(h_{c2}) \propto \delta h^{1/2}$ applies to the all ULS models of phase B, including those near $h_{c1}$ from the right side. Taylor expansion of the aforesaid square root scaling at finite $\delta h = h - h_{c1} > 0$ immediately gives a linear dependence on $\delta h$. The same argument applies to the critical behavior at small $h$ near $h = 0$. In all, near $h = 0$ we have
\begin{equation}
    S_z(h) \propto h^1
\end{equation}
near $h_{c1}$ we have
\begin{equation}
    S_z(h) \propto 
    \begin{cases}
    (h_{c1} - h)^{1/2}, & h < h_{c1}\\
    (h - h_{c1})^1, & h_{c1} < h
    \end{cases}
\end{equation}
and near $h_{c2}$
\begin{equation}
    S_z(h) \propto (h_{c2} - h)^{1/2},\;\;\; h < h_{c2}
\end{equation}
Therefore, in this spinon band language, the two phase transitions at $h_{c1}$ and $h_{c2}$ are both continuous transition in the thermodynamic limit, as a “topological" phase transition of the Lifshitz type that involves 3 distinct spinon bands: the fermi surface (point) of the 1-type spinon vanishes at $h_{c1}$; the fermi surface (point) of the 0-type spinon vanishes and gives rise to the emergence of the (-1)-type at $h_{c2}$.

To further understand the end of the phase diagram shown in Fig.\ref{fig:ulsdyn}(d), we would like to point out that it was obtained in Refs. \cite{PhysRevB.56.14435,PhysRevB.61.4019} that in the spin-1/2 bond operator representation of spin-1, the spin states anti-parallel to the applied field can be projected out, thus the bond operator representation can be approximated by 
\begin{equation} \label{eq:swb}
\begin{split}
        &\frac{S^+}{\sqrt{2}} \sim  u^\dagger t_z \equiv \mathcal{S}^+,\;\; 
        \frac{S^-}{\sqrt{2}} \sim t_z^\dagger u \equiv \mathcal{S}^-,\\
        &S^z \sim \frac{1}{2}(u^\dagger u - t_z^\dagger t_z) + \frac{1}{2} \equiv \mathcal{S}^z + \frac{1}{2}
\end{split}
\end{equation}
where $t_z^\dagger$ creates a triplet state of spin-1/2 bond by $t_z^\dagger \ket{0} = 1/\sqrt{2}(\ket{\uparrow \downarrow} + \ket{\downarrow\uparrow})$ and $u^\dagger$ the bosonic creation operator defined by $u^\dagger\ket{0} = \ket{\uparrow\uparrow}$. Eq.~(\ref{eq:swb}) is the Schwinger boson representation of the pseudo-spin-1/2 operators. Applying such projection produces an effective spin-1/2 anisotropic Heisenberg model subject to an effective magnetic field:
\begin{equation} \label{eq:HeffXXZ}
    H_{eff} \propto  \sum_{\expval{ij}}\mathcal{S}_i^x \mathcal{S}_j^x +\mathcal{S}_i^y \mathcal{S}_j^y + \Delta\mathcal{S}_i^z \mathcal{S}_j^z + h_{eff} \sum_i \mathcal{S}_i^z 
\end{equation}
where $h_{eff} = (h + \beta - 1)/2$ and $\Delta = (1 + \beta)/2$.
This explains the resemblance between the spin-1/2 system and spin-1 system near saturation at $h_{c2}$. Such a mapping from the spin-1 system to the spin-1/2 system is exact for $\beta = 1$ at $h \gtrsim h_{c1}$. In particular, for $\beta = 1$ and $h = h_{c2} = 4$ where ULS model is polarized, the effective field in Eq.(\ref{eq:HeffXXZ}) becomes $h_{eff} = 2$, which is exactly the field that polarizes the spin-1/2 XXZ model \cite{Cabra1998}. At fields above but close to $h_{c1}$, we expect the dynamical structure factor of the ULS model's gapless intermediate phase to coincide with that of the spin-1/2 model calculated by Bethe ansatz, whose intensity decreases as field becomes stronger. 
Figure \ref{fig:ulsall}(c) (for $h = 1.0 > h_{c1})$ validates such a mapping in the intermediate phase of ULS. Moreover, the evolution of dynamical structure factor shows explicitly how the mapping into spin-1/2 model emerges from the bifurcation and recombination of degenerate soft modes.

\section{Discussion}\label{sec:cc}
\subsection{Central Charge}

In this section we turn to a brief discussion of the central charge to provide insight into their effective underlying CFT descriptions
Many pioneering works have been done for the spin-1 antiferromagnetic chain using the NL$\sigma$M \cite{HALDANE1983464,PhysRevLett.50.1153}, which has been recently extended from the Heisenberg model with $\beta = 0$ to the BLBQ Hamiltonian with a wider range of $\beta > 0$ \cite{PhysRevB.102.014447}. 

\begin{table}[H]
\renewcommand{\arraystretch}{1.2}
\begin{tabular}{p{2.8cm}<{\centering}p{1cm}<{\centering}p{2cm}<{\centering}p{2cm}<{\centering}p{2cm}<{\centering}}\toprule
Field & & AKLT & Heisenberg                             \\  \midrule 
                        & $\Delta$ & $>0$ & $>0$                           \\
$h = 0$              & $q_0$    & NA             & NA                    \\
                        & $c$      & NA             & NA                      \\  \midrule 
                        & $\Delta$ & $>0$ & $>0$                       \\
$|h|< h_{c1}$          & $q_0$   & NA             & NA            \\
                        & $c$      & NA             & NA                                 \\  \midrule 
                        & $\Delta$ & $0$             & $0$                                  \\
$h_{c1}<|h|<h_{c2}$ & $q_0$    & $\pi$       & $\pi$                    \\
                        & $c$      & $c\rightarrow 1$   & $c\rightarrow 1 $                      \\   \bottomrule
\end{tabular}
\caption{Summary of energy gap $\Delta$, momentum of gapless modes $q_0$, and central charge $c$ of Heisenberg and AKLT model under different fields. The central charge of the ULS model at fields corresponding to the XXX intermediate phase is $c = 1$ due to its mapping to the spin-1/2 chain. Although such a mapping is no longer exact on the path from $\beta = 1$ to $\beta = 0$, the behavior of the central charge in the B and XXZ phases is shown in Fig.\ref{fig:cc}(d). Away from the phase transition at the corresponding $h_{c1}$ the central charge is close to $c=1$. 
The right arrow in the table $c \rightarrow 1$ indicates that the central charge is unity only away from the transition within the accuracy of the numerical results.}
\end{table}
\begin{figure*}
\centering 
\includegraphics[width=0.99\textwidth]{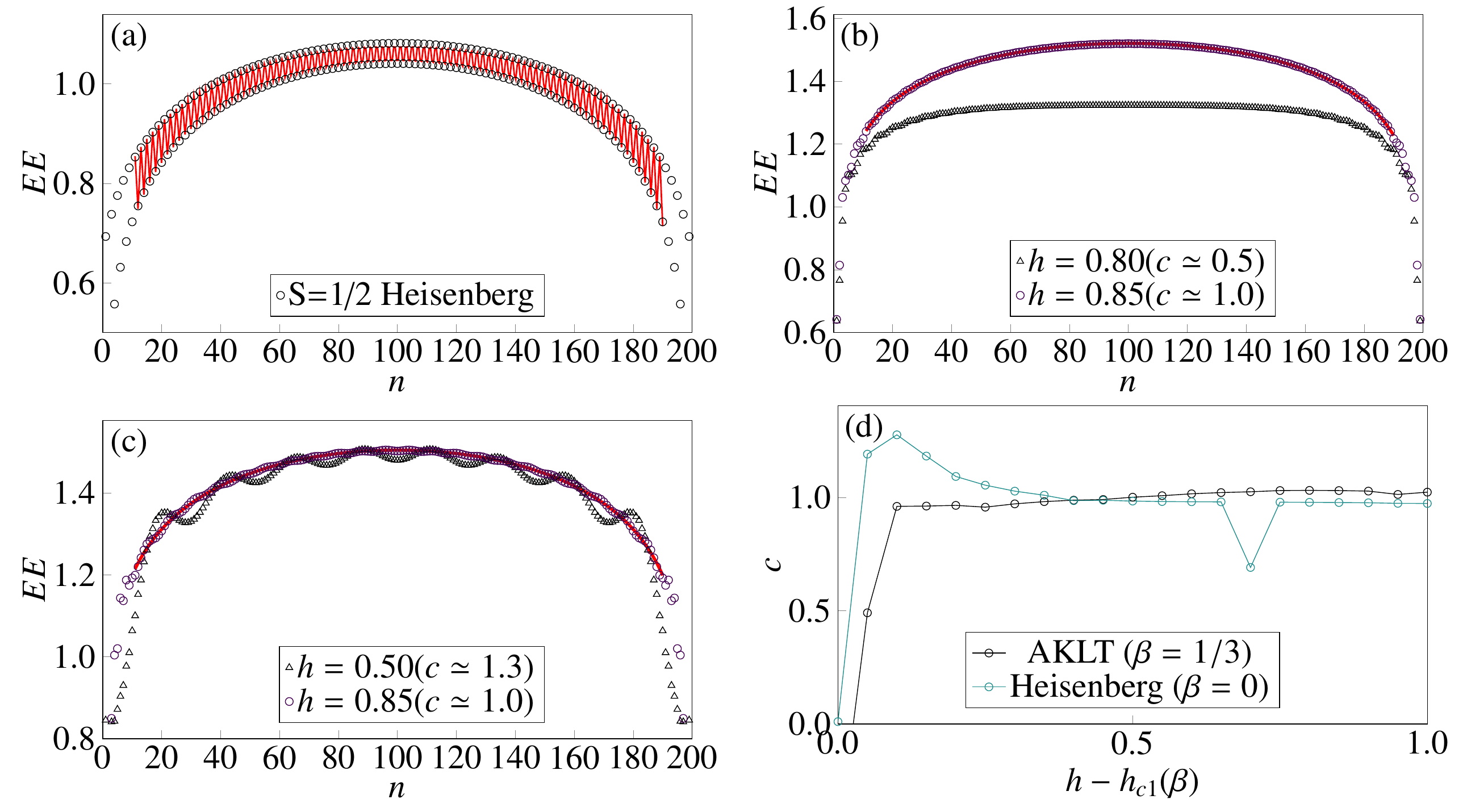}
\caption{\label{fig:cc}   Entanglement entropies as a function of bond position $n$ and fits to the data (red solid line) for (a) the spin-1/2 Heisenberg model with $c = 1$, which is equivalent to ULS under a field larger than $h_{c1}$ (b) AKLT model, (c) spin-1 Heisenberg model. (d) shows central charge $c$ as a function extracted from EE of AKLT and spin-1 Heisenberg model within the intermediate phase. Results obtained from 200-site DMRG under OBC.}
\end{figure*}
While NL$\sigma$M captures the presence of the elementary magnon at $q = \pi$ in the extended region of the Haldane phase, it fails to accurately capture correlation functions beyond the AKLT point at $\beta = 1/3$. In this section, we investigate a possible field theory starting from the Haldane phase boundary at the ULS point at $\beta = 1$ by looking into the central charge of the phase B. We show that perturbation of $\beta$ is irrelevant in phase B, that is, a theory with $c = 1$ is robust for a very wide range of $\beta$ in the gapless regions that emerge in the ULS or Haldane model under a field. 

It is well-known that in the continuum limit an antiferromagnetic spin-1/2 chain can be described by an SU(2) WZW theory with central charge $c = 1$, which can be generalized to many other 1+1-dimensional quantum critical systems with higher SU(N) symmetries with central charge $c = N-1$.  We present below the entanglement entropy at different bonds in models studied in previous sections, and investigate the existence of possible effective CFT correspondence by extracting the central charge in the phase B.

The ULS model at $h=0$ can be captured by a SU(3) WZW theory with $c=2$. Under $h_{c1} < h < h_{c2}$, ULS model can be mapped exactly to a spin-1/2 Heisenberg model \cite{PhysRevB.61.4019,PhysRevB.56.14435}, hence in the low energy regime this
 intermediate phase should have an effective CFT with central charge $c = 1$. As we move away from the ULS point by increasing $\beta$ from $\beta = 1$,  the mapping to the
  spin-1/2 Heisenberg model is no longer exact in the intermediate phase because the anti-parallel states in the bond-operator
  representation cannot be projected out, unless the field is close to saturation at $h_{c2}$. The deviation from the effective spin-1/2 chain can be seen from the difference in their
    spectral weight distribution of intermediate phase shown in Fig.~\ref{fig:akltdyn}(d) and Fig.~\ref{fig:ulsdyn}(d). However, as we are to show next,  the $c = 1$
     theory can be robust against a non-perturbative deviation of $\beta = 1$ where the mapping to the spin-1/2 model is not valid.

The entanglement entropy of 1+1 dimensional CFT under OBC satisfies
\begin{equation} 
    S(n) = S^{CFT}(n) + S^{OSC}(n) + \textrm{const}
\end{equation}
where $n$ is the bond position. The first two terms are as defined in Eq.(\ref{eq:cc}).
Figure.\ref{fig:cc} shows results of entanglement entropy (EE) as a function of bond position $n$, and is fitted by Eq.~(\ref{eq:cc}) to extract the central
 charge at different fields. As a benchmark we show in Fig.~\ref{fig:cc}(a) the EE(n) of the spin-1/2 Heisenberg model with $c = 1$, which is equivalent to ULS model under fields $h_{c1} < h < h_{c2}$. Fig.~\ref{fig:cc}(b)
  shows the EE(n) of the AKLT model near the first critical field $h_{c1}(\beta = 1/3) \approx 0.75$.

It is worth pointing out that while the EE of the spin-1/2 Heisenberg
model oscillates strongly under OBC, the EE of AKLT's intermediate phase does not, hence we can drop the $S_n^{OSC}$ term in Eq.~(\ref{eq:cc}). When 
$\delta h_1 = h - h_{c1} < 0.1$, i.e. $h_{c1} < h < 0.85$, the central charge $c \approx 0.5$ deviates from spin-1/2 chain's $c = 1$ which reflects the 
invalidity of the mapping, yet, as shown in Fig.~\ref{fig:cc}(d), the central charge converges rapidly after $\delta h_1 > 0.1$ to $c = 1$ and remains such up to
saturation at $h_{c2} = 4$. In the spin-1 Heisenberg model where $\beta = 0$, the central charge also quickly converges to $c=1$, but this time EE oscillates with larger amplitude and periodicity than that of AKLT for small $\delta h_1 = h - h_{c1}$ as shown in Fig.~\ref{fig:cc}(c). In the Heisenberg model, the central charge converges to $c = 1$ beyond $\delta h_1 > 0.4$,
i.e. $h > 0.8$ and remains the same until saturation, and the oscillation of EE also disappears after the convergence like that of the AKLT.  Hence, although
$\beta$ deviates non-perturbatively from $\beta = 1$ to AKLT model with $\beta = 1/3$, or to Heisenberg model with $\beta = 0$, a central charge $c = 1$ continues to describe the gapless intermediate phase.

\subsection{Single Mode Approximation} \label{sec:discussion}
Motivated by the sharp signal of the dynamical structure factor in Fig.\ref{fig:akltdyn}(a-d) and Fig.\ref{fig:ulsdyn}(d), we investigate the extent to which a single mode approximation (SMA) can describe the spectrum of the aforementioned models. 

\begin{figure*}[t]
	\includegraphics[width=0.98\textwidth]{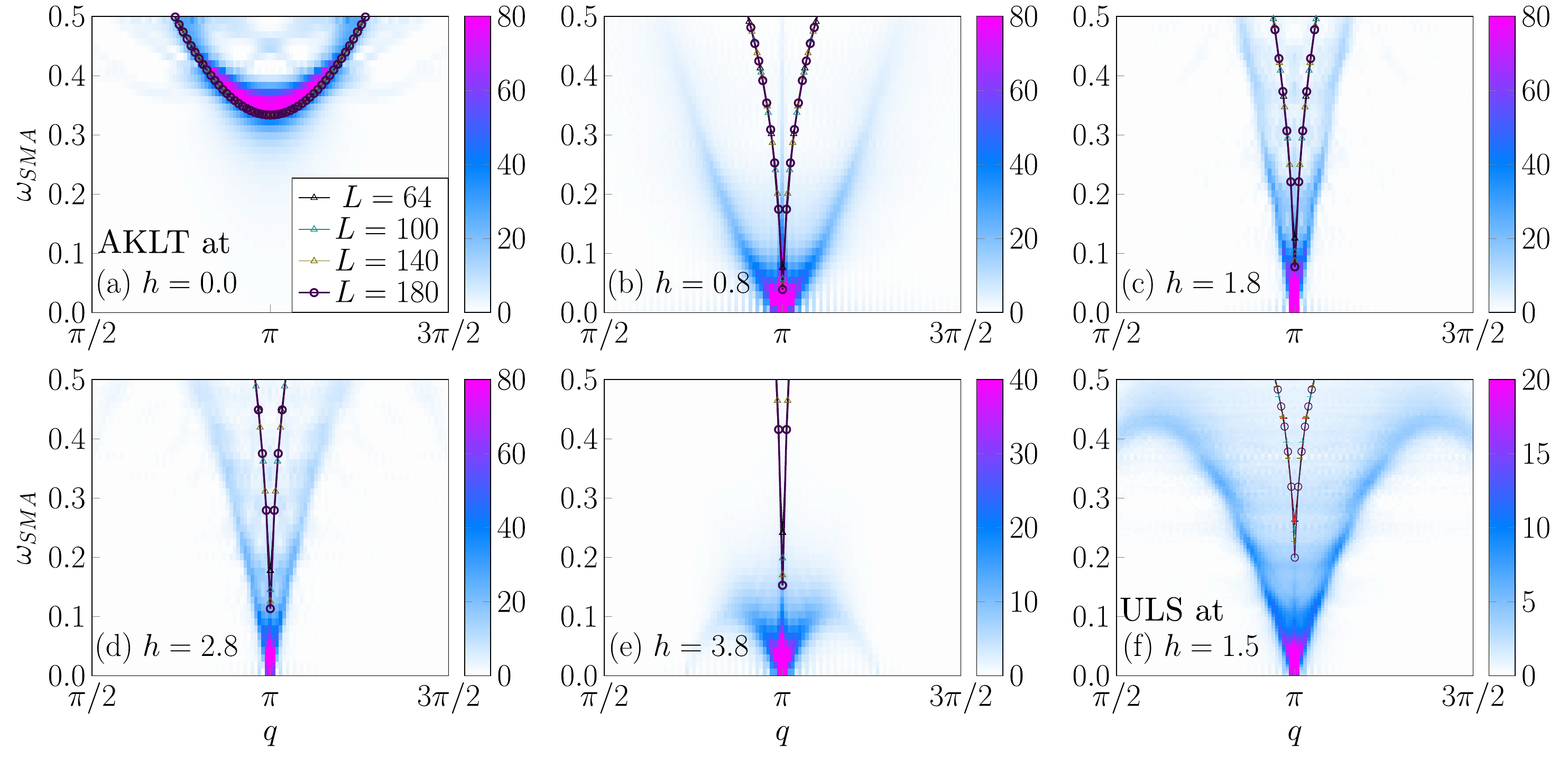}
	\caption{ \label{fig:SMA} SMA results (solid lines) compared with $S(q,\omega)$ obtained by DMRG (intensity plot) in (a) AKLT VBS state; (b-e) AKLT phase B at $h = 0.8$, $h = 1.8$, $h = 2.8$ $h = 3.8 > h_{c1}$ calculated for different $L$. The gap of the VBS state is remarkably close to the exact result of $\Delta_{VBS} = 0.350$. Within the phase B of AKLT, at field slightly larger than $h_{c1}$, e.g. $h = 0.8$ in (a), the (upper bound of) gap given by $\omega_{SMA}(k)$ is close to zero and decreases as $L$ increases, indicative of a gapless mode in the thermodynamic limit. However as $h$ keeps increasing in (b,c), the gapless mode is no longer captured by SMA. (f) ULS: SMA and DMRG for the gapless intermediate phase at $h = 1.5 > h_{c1} \approx 0.94$. The upper bound is not tight enough to capture the gapless mode at $\pi$.}
\end{figure*}
To understand the the nature of the excitations in the intermediate phase B of $H(h;\beta)$ we turn to Bijl and Feynman's SMA method\cite{PhysRev.102.1189}, which successfully described the phonon-roton curve in $^4$He and was later used to explain the antiferromagnetism in extended Heisenberg models with a Haldane gap \cite{PhysRevLett.60.531,PhysRevB.49.15771,Golinelli1999}. SMA assumes the existence of well-defined modes with a sharp dispersion $S(q,\omega) \propto S(q)\delta(\omega - \omega(q))$. 
It was shown previously that SMA works well in capturing the gap above the AKLT ground state \cite{book1}. As we discuss below, SMA is able to capture the essence of the 
field-induced gapless modes at fields above but close to $h_{c1}$ near the Haldane-phase B boundary (see Fig.\ref{fig:phasedig}) with good accuracy.
However, SMA becomes too   
coarse of an approximation to capture the gapless modes at higher fields in the B intermediate phase due to increasing magnon fractionalization.

Though the AKLT chain does not order unless it is in the saturated phase $h>4$, when the requirement for SMA to be valid is rigorously met, nevertheless, we find that SMA can capture the essence of the modes when the fractionalization of the magnon modes is weak, and the deviation from SMA provides a quantitative measure of the degree of fractionalization.
One point to note is that the gap deduced by SMA is the upper bound of the actual gap as is evident from Fig.\ref{fig:phasedig}.  The strength of SMA is that it provides qualitative information on the excitations based only on the static structure factor, without detailed information on the dynamical information.
In order to compare with the well-known result by Arovas, Auerbach and Haldane \cite{PhysRevLett.60.531} we scale down the Hamiltonian by 1/2 hereafter in this section. By SMA we assume the magnon excitations can be described by  
\begin{equation}
	S_q^\alpha\ket{g.s.} = \frac{1}{L}\sum_{i} e^{iqr_i} S_i^\alpha \ket{g.s.}
\end{equation}
where $\ket{g.s.}$ is the ground state of Hamiltonian $H$ which can be readily computed by DMRG, $\alpha=x,y,z$ are three different magnon branches. This is a good approximation if the magnon dispersion is strongly peaked at the energy of the state $S_q^\alpha\ket{g.s.}$. The dispersion within SMA is then given by
\begin{equation} \label{eq:SMA}
	\omega_{SMA} = \frac{\mel{g.s.}{[S_{-q}^\alpha,[H,S_q^\alpha]]}{g.s.}}{2\mel{g.s.}{S_{-q}^\alpha S_q^\alpha}{g.s.}}
\end{equation}
where the denominator is simply the static structure factor evaluated in the ground state. In the presence of inversion symmetry (or PBC) commutators in the numerator can be worked out directly. Here we choose the $S^z$ magnon branch and the energy is evaluated to be:
\begin{equation} \label{eq:SMA2}
	\omega_{SMA} = \frac{(1-\cos q)\mathcal{C}(\beta)}{S(q)}
\end{equation}
where $\mathcal{C}(\beta)$ is a collection of correlators between nearest neighbors. It is also independent of $q$ and fully determined by the choice of the magnon branch, the parameter $\beta$ in the BLBQ Hamiltonian. For a derivation of $\omega_{SMA}$ readers can refer to Appendix \ref{sec:DSMA}. Futhermore, in the $S^z$ branch the $z$-field term of the commutators in Eq.~(\ref{eq:SMA}) vanishes, thus we can simply use the non-perturbed Hamiltonian for the numerator. The spectrum becomes gapless, i.e. $\omega_{SMA}=0$ when the structure factor $S(q)$ diverges, as seen from Eq.(\ref{eq:SMA2}).  

We make no attempt to apply SMA to the phase A 
because it is an extremely fractionalized phase that strongly violates $S(q,\omega) \propto 
 S(q)\delta(\omega - \omega(q))$.  It turns out that, even though for $h_{c1} < h < h_{c2}$ the ULS gives much sharper signal in dynamical structure factors shown in 
 Fig.~\ref{fig:ulsdyn}(d), SMA is a poor approximation to capture the dynamical information, which is reflected by a non-zero minimum of $\omega_{SMA}$ in  Figure.\ref{fig:SMA}(f). This can be qualitatively accounted by the fact that, at least 
 for fields close to but smaller than saturation field $h_{c2} = 4$, the behavior of spin-1 BLBQ chains very much resembles that of a spin-1/2 Heisenberg chain, and can be mapped 
 exactly to spin-1/2 chain for ULS point \cite{PhysRevB.56.14435,PhysRevB.61.4019}.  The dynamical solution of spin-1/2 chain from Bethe ansatz is qualitatively 
 consistent with phase B of ULS shown in Fig.~\ref{fig:ulsdyn}(d), both of which have a $S(q,\omega)$ that resembles a fuzzy fan area of fractionalized signal emanated from $q = \pi$.
 SMA loses too much information by ignoring these fractionalized modes.

Figure.\ref{fig:SMA} shows the single mode dispersion $\omega_{SMA}$ at different fields obtained for different $L$. Fig.~\ref{fig:SMA}(a-e) show results for the AKLT model. In the VBS ground state, the SMA dispersion can be solved exactly: $\omega_{SMA} = 5(5+3\cos k)/27$ with the SMA gap $\omega_{SMA}(\pi) = 0.370$, which is very close to the number $\Delta \approx 0.350$ given by ED. At $h = 0.8$, which is slightly larger than $h_{c1}$ and belongs to the gapless phase B of AKLT, the lowest excitation energy for $L = 180$ obtained by SMA is about $\omega_{SMA} \sim 0.03$ at $q = \pi$, which is tiny compared to that of the VBS state. This upper bound of the gap is affected by the finite size $L$ and decreases with increasing $L$, so we can speculate that there is a gapless mode at fields larger than but close to $h_{c1}$ within the intermediate phase, and the slope of the two nearly linear branches slowly tends to infinity as $h$ gets close to $h_{c1}$, which actually represents a one-dimensional Bose condensation at the critical point \cite{PhysRevB.43.3215}. Such Bose condensation breaks down as the field $h$ increases beyond $h_{c1}$.
 
Figure.\ref{fig:SMA}(b-e) shows the same SMA calculation at larger fields within phase B of AKLT. As the field $h$ increases within B and gets further away from the Haldane-B boundary, the lowest excitation energy captured by SMA no longer converges to zero. This can be readily seen in the static structure factor $S(q)$ in Fig.\ref{fig:akltall}(d), that the spiky $S(q)$ of phase B at $q = \pi$ decreases as magnetic field $h$ increases, hence the approximated gap by $\omega_{SMA}$ in Eq.(\ref{eq:SMA2}) becomes larger. In other words, since the existence of a gapless mode is already guaranteed by the diverging correlation length, a non-zero gap in SMA means the upper bound of the gap is not asymptotically tight, which suggests the assumption $S(q,\omega) \propto S(q)\delta(\omega - \omega(q))$ is no longer accurate for the dynamical structure factor at higher field away from Haldane-B boundary and the Bose condensate breaks down. Further increase of magnetic field enhances fractionalization and ultimately the system resembles a spin-1/2 model somewhere near saturation field $h_{c2}$, where all spin states anti-parallel to the applied field can be asymptotically projected out as in the case of the A-B transition in the ULS model in a field. Therefore, from the calculated data in Fig.\ref{fig:SMA}, we can speculate that the SMA result of AKLT after some large enough field between  Fig.\ref{fig:SMA}(d) and (e) should be the same as an effective XXZ model under an effective magnetic field, whose $S(q,\omega)$ is similar to that of the phase B of ULS shown in Fig.\ref{fig:SMA}(f).  This qualitatively explains the shape of intensity distribution in Fig.\ref{fig:SMA}(e), thus the heavier tail at higher energy in AKLT phase B shown in Fig.\ref{fig:akltall}(k,l), and the decreasing of entanglement entropy at larger fields shown in Fig.\ref{fig:akltall}(c).

\subsection{Material Candidates}
In this subsection we discuss candidate materials with $d^4$ configuration where the field-induced Lifshitz-type transition of ULS Hamiltonian may be observed. 
Contrary to common wisdom that $d^4$ materials are non-magnetic in both strong spin orbital coupling (SOC) and Hund's coupling limits~\cite{balents11-d2},
our recent study using the full multi-orbital Hubbard model of $d^4$ configurations indicates that a magnetic phase transition is possible in a realistic parameter regime~\cite{PhysRevB.91.054412,Trivedi1_d4}. 
For example, Ca$_2$RuO$_4$ was shown to have finite 
magnetic moments~\cite{CaRuO_1,CaRuO_2,CaRuO_3}, and 
experiments on double perovskite iridates ~\cite{PeroSkite_1,PeroSkite_2,PeroSkite_3,BaYIrO_1,BaYIrO_2,BaYIrO_3,BaYIrO_4}, honeycomb ruthenates \cite{Ruthnate_1,Ruthnate_2,Ruthnate_3} have also revealed non-trivial magnetism for the $d^4$ configuration. 

In our previous work~\cite{PhysRevB.101.155112} using DMRG on the model derived for $d^4$ transition metal oxides we also find a gapless-to-gapless transition with increasing SOC. The behavior near the transition point of the $d^4$ model captured by a mean-field theory described by a ULS model comprised of only $L=1$ orbital degrees of freedom at $\beta = 1$ with an additional spin-orbital interaction.  In the following we briefly describe the origin of the model and its connection with the ULS Hamiltonian in the orbital sector.

The effective Hamiltonian for $d^4$ materials is effectively described by \cite{PhysRevB.91.054412,PhysRevB.101.155112}:
\begin{equation}\label{eq:hamiltonian_d41}
    \begin{split}
      H_{d4} = &-J \sum_{\< ij \> }(\mathbf{S}_i \cdot \mathbf{S}_j ) \mathcal{P}(\mathbf{L}_i + \mathbf{L}_j = 1) \\ 
      &+ \lambda \sum_i  \mathbf{L}_i \cdot \mathbf{S}_i,        
    \end{split}    
\end{equation}
The effective coupling constants $J$ and $\lambda$ represent ferromagnetic exchange and spin-orbit interactions. The projection operator in the first term, $\mathcal{P}(\mathbf{L}_i + \mathbf{L}_j = 1)=- \frac{1}{8} \mathbf{L}_{\<ij\>}^2 (\mathbf{L}_{\<ij\>}^2 - 6)$, is defined on a bond connecting orbital sectors of two adjacent sites. For a two site problem, the total orbital angular momentum can be $L_T = 0,1$ or $2$. Therefore, the projector $\mathcal{P}(\mathbf{L}_i + \mathbf{L}_j = 1) = 0, 1, 0$ for $L_T = 0,1,2$ respectively. 
For $J > 0$, this  projector makes the $L_{\<i j\>} = 0$ and $2$ quantum sectors energetically unfavorable on the two-site bond, while preferring $L_{\<i j\>} = 1$ angular momentum on the bond. Upon expanding the projector we arrive at the explicit form of the Hamiltonian:  
\begin{equation}\label{eq:hamiltonian_d4}
    \begin{split}
          H_{d4} = &\frac{J}{2} \sum_{\< ij \> }(\mathbf{S}_i \cdot \mathbf{S}_j ) \left( (\mathbf{L}_i \cdot \mathbf{L}_j)^2 + \mathbf{L}_i \cdot \mathbf{L}_j - 2 \right) \\
     & + \lambda \sum_i \mathbf{L}_i \cdot \mathbf{S}_i, 
    \end{split}
\end{equation}
 In Ref.\cite{PhysRevB.101.155112} we have shown that the model exhibits an emergent spin-orbital separation in a spin-orbital interacting system. Therefore we can factorize our Hamiltonian into a spin and orbital part, similar to a mean-field approximation, 
\begin{equation} \label{eq:mf_hamiltonian}
    \begin{split}
       H_{d4} \approx &\frac{J}{M_S^2}{2} \sum_{\langle ij\rangle } \Bigl( (\mathbf{L}_i \cdot \mathbf{L}_j)^2 + \mathbf{L}_i \cdot \mathbf{L}_j -2 \Bigr) 
        \\ & + \lambda M_S \sum_{i} L_i^z  
    \end{split}
\end{equation}
%
where we assume $M_s^2 \simeq \sum_{\<ij\>} \< \mathbf{S}_i \cdot \mathbf{S}_j \>$ and treat the spin-orbit coupling in the Ising limit with $M_S = \sum_i |\<S_i^z\>|$. This approximation is justified by the numerical results that shows the magnetization of spins remains large for weak enough SOC. 
In summary, the similarity with Ref.~\cite{PhysRevB.58.R14709} lead us to expect that our model (Eq.~\ref{eq:hamiltonian_d41}) can be well approximated by only the orbital term in the $H_{d4}$ Hamiltonian with a Zeeman field $L^z$ (Eq.~\ref{eq:mf_hamiltonian}). The first term in $H^{mf}$ is exactly the ULS Hamiltonian up to a constant. Therefore the effective Hamiltonian can be interpreted as the ULS Hamiltonian with an additional Zeeman field. Setting the energy scale $J = 1$ we have:
%
\begin{align}\label{eq:ULSZ_hamiltonian}
  H_{eff} =& \sum_{\langle ij\rangle } \Bigl( (\mathbf{L_i} \cdot \mathbf{L_j})^2 + \mathbf{L_i} \cdot \mathbf{L_j} \Bigr) + h_{eff} \sum_i L_i^z ,
\end{align}
%
where $\mathbf{L}_i$ are the spin-1 Pauli operators at site $i$, and $h_{eff} = 2\lambda/M_s$ is the strength of an effective external Zeeman field experienced by the orbital degree of freedom. This is an orbital analog of ULS model with a field, as discussed in Sec.\ref{sec:uls}  Eq.(\ref{eq:ulsfermion1}). Therefore we expect our predictions in Sec.\ref{sec:uls} are useful to guide explorations for $5d^4$ transition metal oxides like OsCl${}_4$, Ca$_2$RuO$_4$, and other double perovskite iridates.
 
\section{Summary and Outlook} \label{sec:summary}
In summary, we have investigated the one-parameter bilinear biquadratic Hamiltonian family for two parameter values $\beta = 1/3$, the AKLT model as a representative of the Haldane phase, which is compared with $\beta = 1$, the ULS critical point and shown the process by which the ground state evolves under a magnetic field. Both models undergo two second order transitions: the AKLT model first transitions from the Haldane phase to a gapless phase B, and then to the fully saturated phase at field $h_{c2} = 4$. The ULS critical point, already gapless at zero field, goes through a second transition to the gapless phase B before reaching the fully polarized phase at $h_{c2} = 4$. We showed that the gapless to gapless transition in ULS model under a field can be understood as a Lifshitz type transition that involves 3 distinct spinon bands
in the gapless to gapless transition. 
In the spinon band language, the two phase transitions at $h_{c1}$ and $h_{c2}$ are both continuous transition in the thermodynamic limit, as a “topological" phase transition of the Lifshitz type that involves 3 distinct spinon bands: the fermi surface (point) of the 1-type spinon vanishes at $h_{c1}$; the fermi surface (point) of the 0-type spinon vanishes and gives rise to the emergence of the (-1)-type at $h_{c2}$. 
We have scrutinized the universality of central charge in the gapless phase B which can be effectively captured by a CFT with $c=1$. We expect our predictions of the spin dynamics will open the door for inelastic neutron scattering measurements in candidate materials of relevant quasi-one dimensional $d^4$ materials \cite{PhysRevB.91.054412,PhysRevB.101.155112}.
Future theoretical work will involve the nature of edge modes of BLBQ models under OBC, the effect of thermal fluctuations on symmetry protected topological states in addition to a field, and a field-theoretic  approach to determine the effective CFT to describe the gapless intermedate phases. 

\section{acknowledgements}
We thank Dr. E. Miles Stoudenmire for help with the Intelligent Tensor Library (ITensor) open source code.
Most of the results for the static results were obtained with ITensor \cite{fishman2020itensor};
the dynamics (real frequency) results were obtained with \textsc{DMRG++} \cite{re:alvarez09}, and see also \cite{re:supplemental}.
S.F. and N.T. acknowledge support from DOE grant DE-FG02-07ER46423. Computations were performed using the Unity cluster at the Ohio State University and Ohio super computing center (OSC).
G.A. was supported by the Scientific Discovery through Advanced Computing (SciDAC) program
funded  by  U~S.~Department of Energy,  Office of Science, Advanced Scientific Computing Research and Basic Energy Sciences, Division of Materials Sciences and Engineering.
GA was also supported by the ExaTN ORNL LDRD.

\section{Appendix}

\subsection{Derivation of $\mathcal{C}(\beta)$ of SMA} \label{sec:DSMA}
In this section we sketch the derivation of $\mathcal{C}(\beta)$ mentioned in Eq.~(\ref{eq:SMA2}). As an example we will derive the $S_zS_z$ channel. Here we use the conventional BLBQ parameterized by $\beta$:
\begin{equation}
	H = \sum_{i,i+1} \left( \vec{S}_i\cdot \vec{S}_{i+1} \right) + \beta \left( \vec{S}_i\cdot \vec{S}_{i+1} \right)^2
\end{equation}
The evaluation of SMA can be reduced to the evaluation of commutator in Eq.~(\ref{eq:SMA}). 
\begin{widetext}
\begin{equation}
	\begin{split}
		\left[ S_{-q}^\alpha , [H_\beta, S_q^\alpha] \right]
        &= \sum_{inn'}  \left( \left[S_{n'}^\alpha,  \vec{S}_i\cdot \vec{S}_{i+1}\right]S_n^\alpha - S_{n}^\alpha \left[ S_{n'}^\alpha, \vec{S}_{i}\cdot \vec{S}_{i+1} \right] \right)\frac{e^{-iq(n-n')}}{L} \\
								      &- \beta \left( \left[ S_{n'}^\alpha, \left( \vec{S}_i\cdot \vec{S}_{i+1} \right)^2 \right]S_n^\alpha - S_n^\alpha \left[ S_{n'}^\alpha, \left( \vec{S}_i\cdot \vec{S}_{i+1} \right)^2 \right] \right)\frac{e^{-iq(n-n')}}{L} 
								      \equiv H_L + H_Q
	\end{split}
\end{equation}
\end{widetext}
where we have moved to Fourier basis. To two terms in above equation have are from the linear and quadratic term of BLBQ Hamiltonian respectively. We evaluate the linear term first then the quadratic term.  The linear term in the commutator is
\begin{equation}
	\sum_{inn'} \left(\left[S_{n'}^z,  \vec{S}_i\cdot \vec{S}_{i+1}\right]S_n^z  - S_{n}^z \left[ S_{n'}^z, \vec{S}_{i}\cdot \vec{S}_{i+1} \right] \right)e^{-iq(n-n')}
\end{equation}
Noting that if $n \neq i$ and  $n \neq i+1$, then the commutation factor of linear term must vanish. So we only need to add up these two indices. In presence of inversion symmetry the summation becomes, this gives
\begin{equation}
             H_{L}		
			 = -2(1 - \cos q) \frac{1}{L} \sum_{i} S_i^y S_{i+1}^y + S_i^x S_{i+1}^x 
\end{equation}
where the normalized sum over $i$ contributes to $\mathcal{C}(\beta)$. This alone will be the SMA for the spin-1 Heisenberg chain. Note that the two point correlation in the summation will evaluate to a negative number in antiferromagnetic chain, so the dynamical signal is proportional to $1 - \cos q$. In a similar way we apply the inversion symmetry to $H_Q$ and have
which is again proportional to $1 - \cos q$. So, in arbitrary units we may ignore the multiple-point correlation functions that are independent of momentum during the evaluation of single-mode dynamical structure factor.  In fact for AKLT model $\mathcal{C}(\beta) \sim 1$~\cite{Golinelli1999} so the arbitrary units should be very close to the actual value.

\subsection{Fermionization of ULS Model} \label{sec:ulsfermion}
In this section we show a detailed derivation of ULS's fermionic representation. Rewriting the Hamiltonian in the fermion language is allows us readily notice the SU(3) symmetry as mentioned in the main text, and is a useful tool in finding conserved charges that is not explicit otherwise. For spin-1 sites, define spin-1 operator $\textbf{S}_i = \psi_i^\dagger \vec{S} \psi_i$, where $\vec{S}$ is spin matrix in spin-1 Hilbert space:
\begin{equation}
    S^z = \begin{pmatrix}
        1 \;&  0 & 0 \\
        0 \;&  0 & 0 \\
        0 \;&  0 & -1
    \end{pmatrix}, \;
    S^{+} = 
    \begin{pmatrix}
            0 \;&  \sqrt{2} & 0 \\
            0 \;&  0 & \sqrt{2} \\
            0 \;&  0 & 0
    \end{pmatrix}
\end{equation}
and $\psi = (a_{i,1},a_{i,0},a_{i,-1})^T$, with $a_{i,m} (a_{i,m}^\dagger)$ being fermion annihilation (creation) operator of spin-m at site $i$. So we have
\begin{align}
    S^z &= a_1^\dagger a_1 - a_{-1}^\dagger a_{-1}\\
    S^+ &= \left( S^- \right)^\dagger = \sqrt{2} \left( a_1^\dagger a_0 + a_0^\dagger a_{-1} \right)
\end{align}
There is a constraint that the spin on each site is $1$, thus
\begin{equation}
	\begin{split}
		\frac{1}{2}\textbf{S}^2 &= \frac{1}{2}S^zS^z + \frac{1}{4} \left( S^+ S^- + S^- S^+ \right)\\
			     &=  (n - n_0 n_1 - n_0 n_{-1} - n_1 n_{-1}) 
			     \stackrel{!}{=} 1
	\end{split}
\end{equation}
where $n_m$ is the \emph{on-site} occupation number operator of $m$-type fermion, and $n_i = \sum_{m}a^\dagger_{im} a_{im}$ being the total \emph{on-site} occupation number operator.
The standard ULS Hamiltonian in spin language is written as
\begin{equation}
	H_{ULS} = \sum_{\expval{ij}}\textbf{S}_i \cdot \textbf{S}_j + \left( \textbf{S}_i\cdot \textbf{S}_j \right)^2 - 2I
\end{equation}
note the identity $I$ is spanned in $3\otimes 3$ Hilbert space. The on-site identity is $I_0 = \sum_{\alpha}\mel{\alpha}{I_0}{\alpha}a_\alpha^\dagger a_\alpha = \sum_{\alpha}a_\alpha^\dagger a_\alpha = n  $, hence $I = n_i n_j$. Therefore we can make use of the fermion representation of identity as auxiliary parameters. Let us define a diagonal constant $c = n_i n_j + 3n_i$, then the equivalent ULS Hamiltonian $\mathcal{H}$ can be expressed by.
\begin{widetext}
\begin{equation}
    \begin{split}
            \mathcal{H}_{ULS} &= H_{ULS} - c =   \sum_{\expval{ij}} \mathbf{S}_i \cdot \mathbf{S}_j + ( \mathbf{S}_i \cdot \mathbf{S}_j)^2 - c\\
            &=  - \Big[\sum_{\expval{ij}} a_{i,1}^\dagger a_{j,1} a_{j,1}^\dagger a_{i,1} +  a_{i,0}^\dagger a_{j,0} a_{j,1}^\dagger a_{i,0} + a_{i,-1}^\dagger a_{j,-1} a_{j,-1}^\dagger a_{i,-1} + a_{i,1}^\dagger a_{j,0} a_{j,0}^\dagger a_{i,1} + a_{i,0}^\dagger a_{j,-1} a_{j,-1}^\dagger a_{i,0} \\ 
            &+\sum_{\expval{ij}}  a_{i,1}^\dagger a_{j,-1} a_{j,-1}^\dagger a_{i,1} + a_{i,0}^\dagger a_{j,1} a_{j,1}^\dagger a_{i,0} + a_{i,-1}^\dagger a_{j,0} a_{j,0}^\dagger a_{i,-1} + a_{i,-1}^\dagger a_{j,1} a_{j,1}^\dagger a_{i,-1} \Big]\\
            &= - \sum_{\expval{ij}; mm'} a_{i,m}^\dagger a_{j,m} a_{j,m'}^\dagger a_{i,m'} 
            = -\sum_{\expval{ij}} (\psi_i^\dagger \psi_j) (\psi_j^\dagger \psi_i)
    \end{split}
\end{equation}
\end{widetext}
It is then obvious that $\mathcal{H}_{ULS}$ remains invariant under transformations in $SU(3) \equiv \{ U \in GL(3,\mathbb{C})| U^\dagger U = \mathbb{I},\; \det(U) = 1\}$.
It is then straightforward to show the 3 conserved quantities explicitly by the fermion representation. 
That is, $[N_n, H_{ULS}] = \left[\sum_{n} a_n^\dagger a_n, H_{ULS}\right] = 0$. Hence the total occupation number $N_m$ of $m = -1, 0, 1$ are good quantum numbers separately.
\bibliography{reference.bib}
\end{document}


\title{Supplemental material for “Gapless to gapless phase transitions in quantum spin chains"}

\author{Shi Feng}
\affiliation{Department of Physics, The Ohio State University, Columbus, Ohio 43210, USA}
\author{Gonzalo Alvarez}
\affiliation{Computational Sciences and Engineering Division and Center for Nanophase Materials Sciences, Oak Ridge National Laboratory, Oak Ridge, Tennessee 37831, USA }
\author{Nandini Trivedi}
\affiliation{Department of Physics, The Ohio State University, Columbus, Ohio 43210, USA}
\maketitle

This supplemental contains the following sections. Section I analyzes
the convergence of our results with both the number of DMRG kept states $m$ and
with respect to the thermodynamic limit.
Section II contains additional dynamics figures that are too similar to the ones
presented in the paper, or that did not fit
in the paper. Section III describes the contents of the data sources packaged in \verb!sourcesPaper103.tar.xz! provided
with this supplemental.  Section IV explains how
to obtain, configure, compile, link, and run the \textsc{DMRG++} computer program
to reproduce the dynamical results. We have also added a description of
the inputs provided in \verb!inputsPaper103.tar.xz! that are helpful in obtaining the
dynamical figures. 
Section V presents the evaluation of single mode approximation
of the bilinear biquadratic model.
Section VI presents the detailed derivation of ULS's fermionization, and shows explicitly the SU(3) symmetry and conserved quantum numbers.
\emph{This supplemental and supporting files is also available at}
\nolinkurl{https://g1257.github.io/papers/103} and \nolinkurl{https://zenodo.org/record/4310066}.

\section{Size Effects and convergence}

\begin{figure}[h]
	\includegraphics[width=0.49\textwidth]{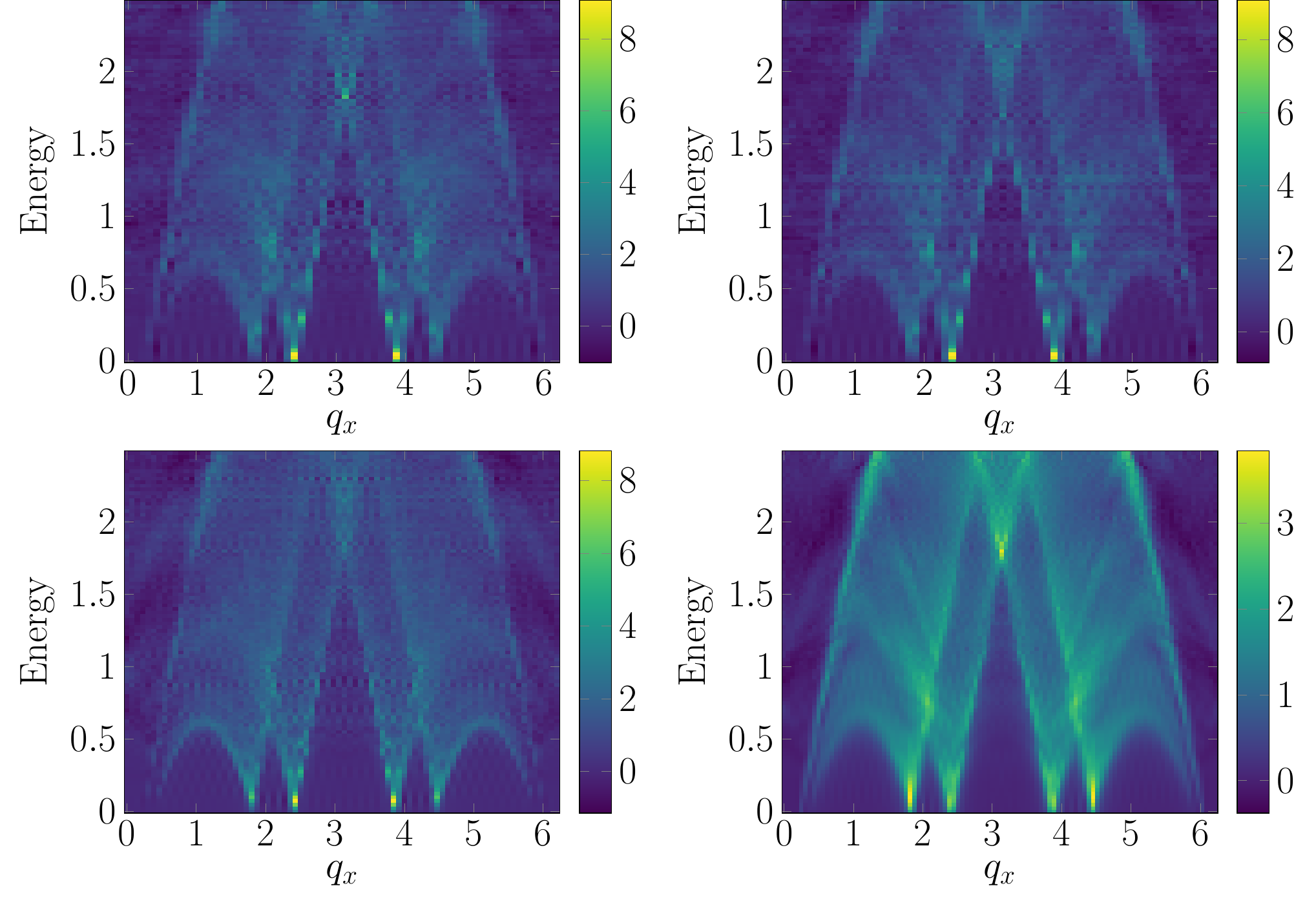}
	\caption{$S^+S^-$ dynamics for the ULS model at zero field for
Above, left: $L=60$ sites, $m=500$ kept states;
Above, right: $L=60$ sites, $m=1000$ kept states;
Below, left: $L=80$ sites, $m=500$ kept states;
Below, right: $L=100$ sites, $m=500$ kept states.}
\label{fig:convergence}
\end{figure}

This section provides evidence that the results presented in the paper are converged.
We analyze convergence first with the DMRG parameter $m$, the number of kept states.
Figure \ref{fig:convergence} shows a comparison of the $S^+S^-$ dynamics for the ULS
model at zero field and with $L=60$ sites for (a) $m=800$ and (b $m=100$ states,
indicating that none of the details discussed in the paper is affected by 
doubling the number of kept states, so that we have worked with correct accuracy
for the level of detail needed to make the conclusions reached in the paper.

Figure \ref{fig:convergence} also shows the effect of increasing the number of sites
from (a) $L=60$ to (c) $L=80$ to (d) $L=100$ while maintaining a constant $m=800$.
Again, no relevant changes are noticeable that would change the conclusions
reached in the paper.

Note that although $m = 800$ is enough for the convergence of dynamical structure factors, it turned out that, for the ULS critical point, we need significantly larger amount of kept states to produce an accurate von Neumann entanglement entropy $S(n)$ that is able to capture the correct central charge. In order to get the central charge for ULS model, we kept $m = 3000$ states which produces the expected $c = 2$ for the SU(3) WZW theory. 
In contrast, smaller $m < 2000$ will give a fitted central charge $c < 2$ though it captures other dynamic/static features with good accuracy. This discrepancy of kept states between central charge calculation and other dynamical/static signature is highly likely to originate from the fine figures of oscillation term $S^{OSC}(n)$ under OBC, which is defined in Eq.(9,10) of the main text, though the envelope from $S^{CFT}$ only needs a much smaller $m$.

\section{Additional Dynamics Data}
This section contains additional dynamics figures that did not fit in the paper.
Figure \ref{fig:dos60zz} shows the density of states $S^zS^z$ for the AKLT model,
and computed with the DMRG on a 60-site lattice with OBC, 
in different spins sectors, as indicated. 
Figure \ref{fig:szszDynamicsAklt} shows the $S^zS^z$ dynamics for the AKLT model
on a $L=60$ site lattice, for the magnetic fields indicated.
We compute and show the $S^zS^z$ dynamics separate from the $S^+S^-$ dynamics,
because they differ in the presence of a magnetic field. As explained in the main text,  the $S^zS^z$ dynamics is similar to that at $h=0$, because, (i) the ground state remains a VBS state, and,
(ii) $H-E_0$ does not depend on field, as the energy contribution of field in $H$ and $E_0$ cancels out.

\begin{widetext}
\begin{minipage}{0.46\textwidth}
\begin{figure}[H]
\includegraphics[width=\textwidth]{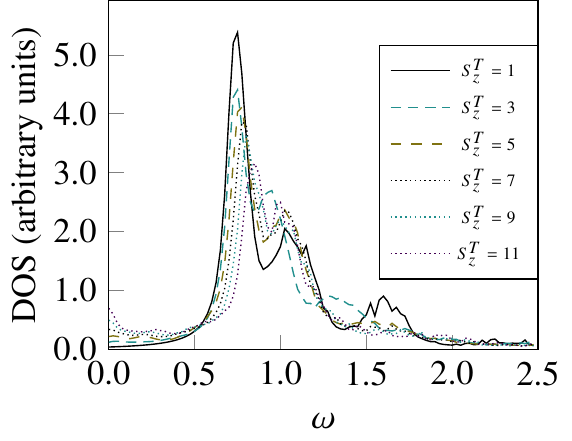}
\caption{DOS for the AKLT model, as defined in the text and computed with DMRG
on a 60-site lattice with OBC,
in different spin sectors, as indicated.}
\label{fig:dos60zz}
\end{figure}
\end{minipage}
\begin{minipage}{0.1\textwidth}

\end{minipage}
\begin{minipage}{0.49\textwidth}
\begin{figure}[H]
\includegraphics[width=\textwidth]{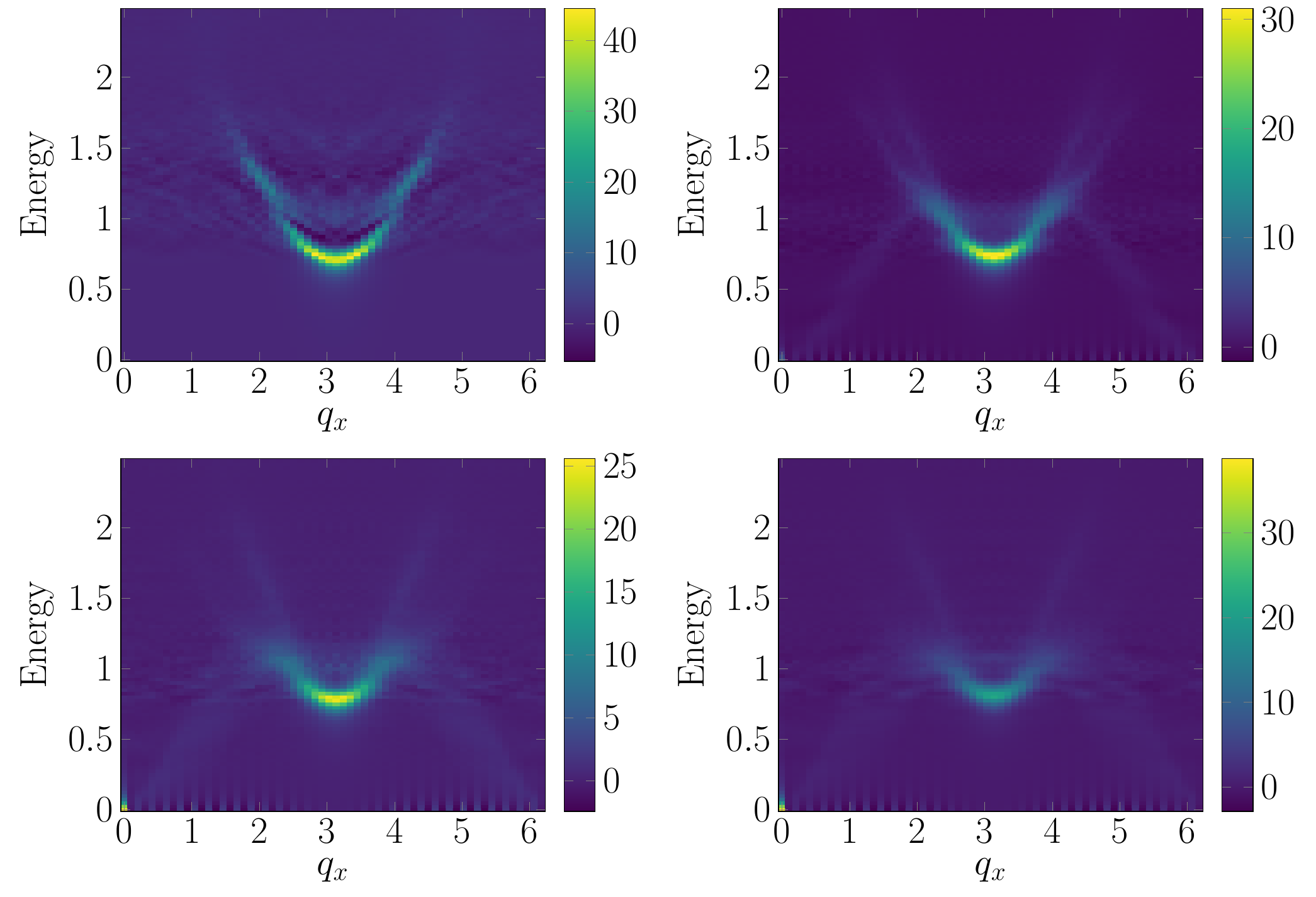}
\caption{\label{fig:szszDynamicsAklt}First row: $S^{zz}(q,\omega)$ at field $h_z=0$ ($sz=1/60$),
and at field $h_z=0.5$ ($sz \approx 0.16$).
Last row:  $S^{z z}(q,\omega)$  at field $h_z=1.0$ ($sz \approx 0.50$), and at field $h_z=1.5$ ($sz \approx 0.60$).
This is a 60-site DMRG run under OBC for the AKLT model. 
}
\end{figure}
\end{minipage}\vspace*{0.5cm}

\begin{minipage}{0.46\textwidth}
\begin{figure}[H]
	\includegraphics[width=\textwidth]{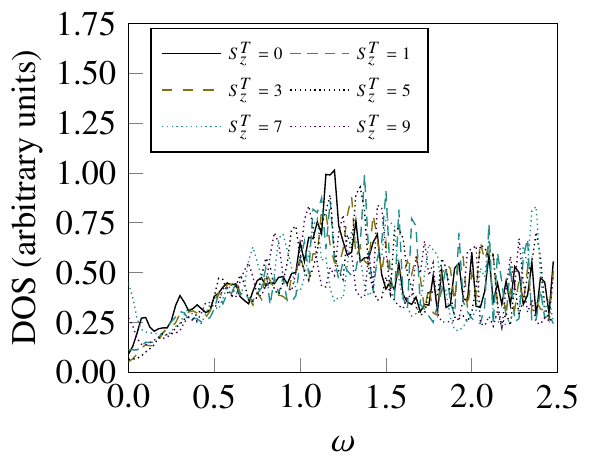}
	\caption{DOS for the ULS model, as defined in the text and computed with DMRG, 
		in different spin sectors as indicated, on a 60-site lattice with
		OBC.}
	\label{fig:dos60zzULS}
\end{figure}
\end{minipage}	
\begin{minipage}{0.1\textwidth}

\end{minipage}
\begin{minipage}{0.49\textwidth}
\begin{figure}[H]
	\includegraphics[width=\textwidth]{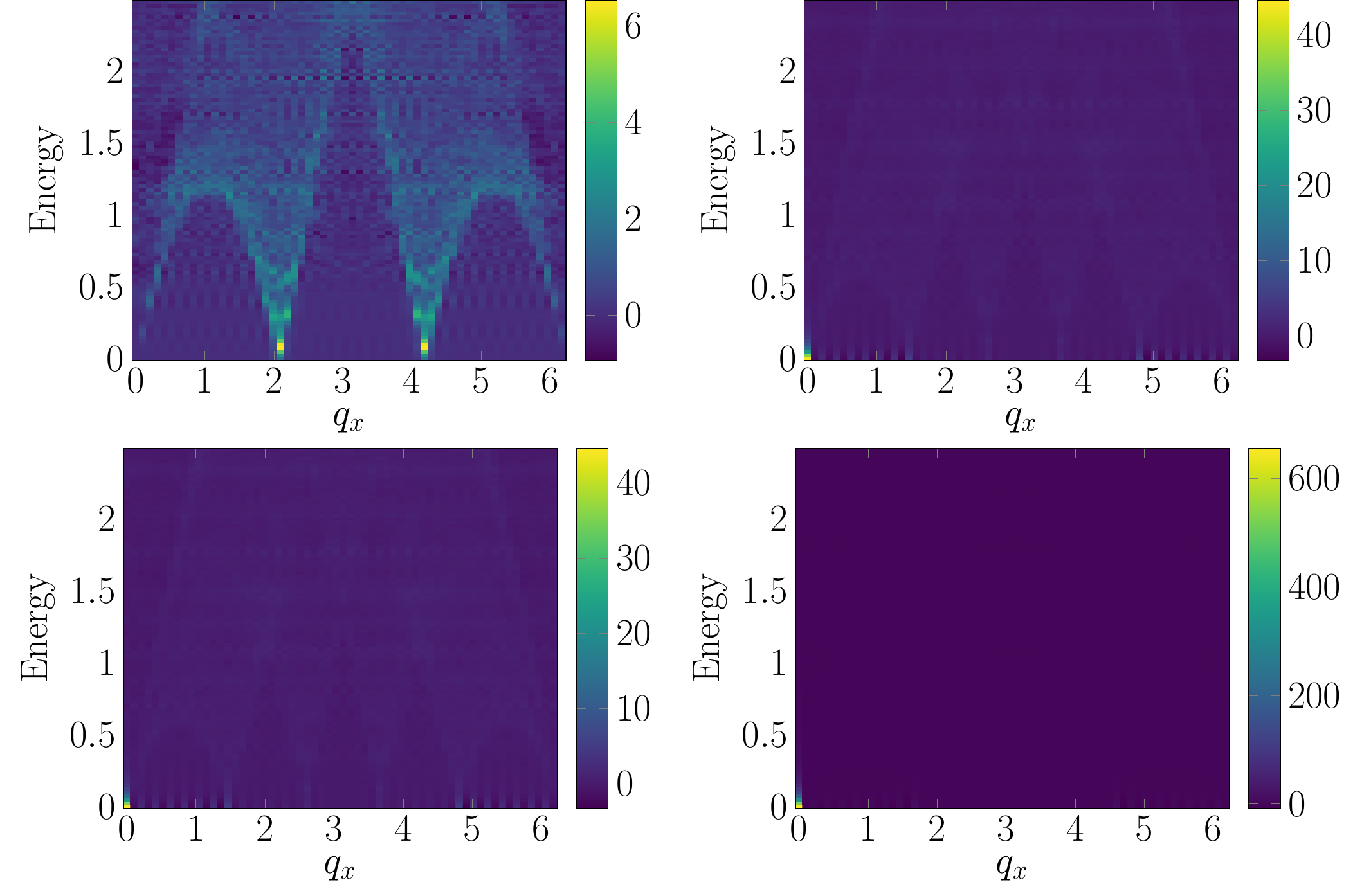}
	\caption{\label{fig:ulsSzszDynamics}First row: $S^{zz}(q,\omega)$ at field $h_z=0$
		and at field $h_z=0.5$.
		Last row:  $S^{zz}(q,\omega)$  at field $h_z=1.5$, and at field $h_z=3$.
		This is a 60-site DMRG run under OBC for the ULS model.}
\end{figure}
\end{minipage}
\vspace*{0.5cm}
\begin{figure}[h]
	\includegraphics[width=\textwidth]{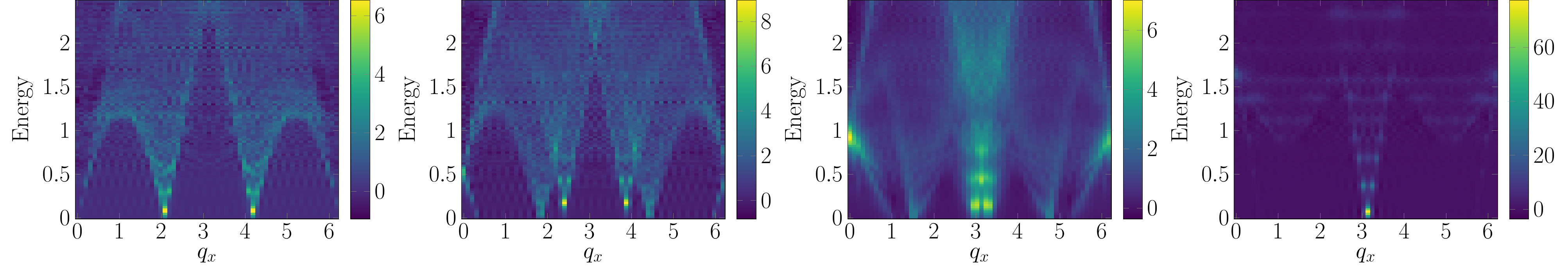}
\caption{\label{fig:ulsSmSpDynamics}First row: $S^{- +}(q,\omega)$ at field $h_z=0$ ($sz = 1/60$),
	and at field $h_z=0.5$ ($sz \approx 0.16$).
	Last row:  $S^{- +}(q,\omega)$  at field $h_z=0.9$ ($sz \approx 0.50$), and at field $h_z=1.5$ ($sz \approx 0.60$).
	This is a 60-site DMRG run under OBC for the ULS model.}
\end{figure}

\begin{figure}
    \centering
    \includegraphics[width=\textwidth]{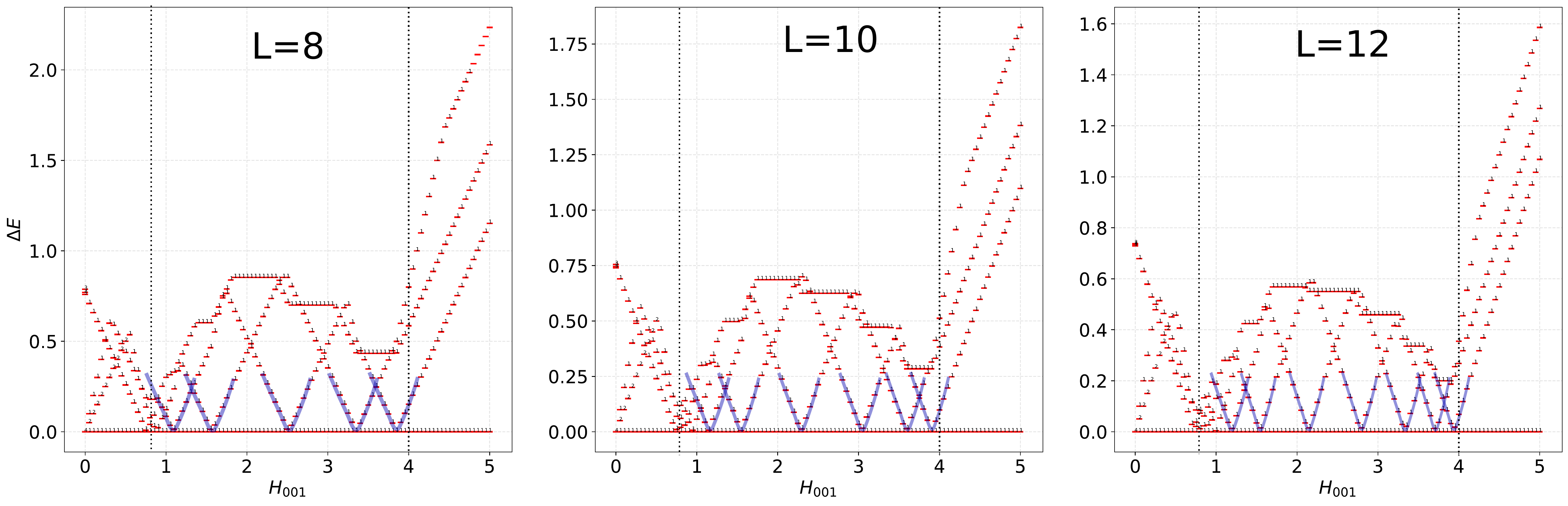}
    \caption{Energy spectrum (measured against g.s. energy $E_0$) of AKLT model with $L = 8, 10, 12$ obtained by ED. Two vertical dashed lines mark the Haldane-B transition at $h_{c_1}$ and B-saturated transition at $h_{c_2} = 4$. Within the intermediate phase $h_{c_1} < h < h_{c_2} = 4.00$, the number of level crossings rises as system size grows. It is expected to fill the entire phase, hence a gapless phase with power-law decaying of correlation function. The blue curves are to guide the eye.}
    \label{fig:Espectrum}
\end{figure}
\end{widetext}
Figure \ref{fig:dos60zzULS} shows the DOS of the ULS model for the parameters indicated.
Figure \ref{fig:ulsSzszDynamics} shows the $S^zS^z$ dynamics for the ULS model, on
a $L=60$ lattice for the magnetic fields indicated in the caption of the figure.

Figure \ref{fig:ulsSmSpDynamics} shows the $S^-S^+$ dynamics for the ULS model, on
a $L=60$ lattice for the magnetic fields indicated in the caption of the figure.
This dynamical structure factor component differs from the $S^+S^-$ one at non-zero magnetic field.

Finally, in order to have an intuition of the Haldane-B transition we show in Fig.\ref{fig:Espectrum} the energy spectrum of AKLT model with $L = 8, 10, 12$ obtained by ED. In the small system, there are successively level crossings in the intermediate phase (marked by blue lines), the number of which increases as the system size $L$ becomes larger. One may perceive the gapless phase in a large finite system as a region filled up with these crossings.

\subsection{Spin-1 Heisenberg Model} \label{sec:heisenberg}
As a benchmark model here we show the results of spin-1 Heisenberg model (Haldane Model). It corresponds to the parameter $\beta=0$ of the  BLBQ family and it also lies in the Haldane phase.  
\begin{equation} \label{eq:ulsfermion3}
        H_{ULSZ} =  \sum_{\expval{ij}} \mathbf{S}_i \cdot \mathbf{S}_j  + h \sum_i S_i^z
\end{equation}
The Heisenberg model's  SU(2) symmetry
breaks down when applying a magnetic field $h$ in the z-direction.
Its ground state with $h$ field becomes then the ground state of the block with some total $S_z$ of the model without field.
In the spin-1 Heisenberg model, there is a second-order phase transition at $h_{c1}\approx0.5$ in units
of energy. As explained in the appendix of maintext, the value $h_{c2}=4$ is the point where the fully saturated phase is reached. The behavior of von-Neumann entanglement entropy and magnetization are similar to that of AKLT except at a different $h_{c1}$,  where the haldane gap protects the original ground state until the field reaches the energy scale at which the gap closes and a level crossing made the old excited state the new ground state. 

\begin{figure}
	\includegraphics[width=0.49\textwidth]{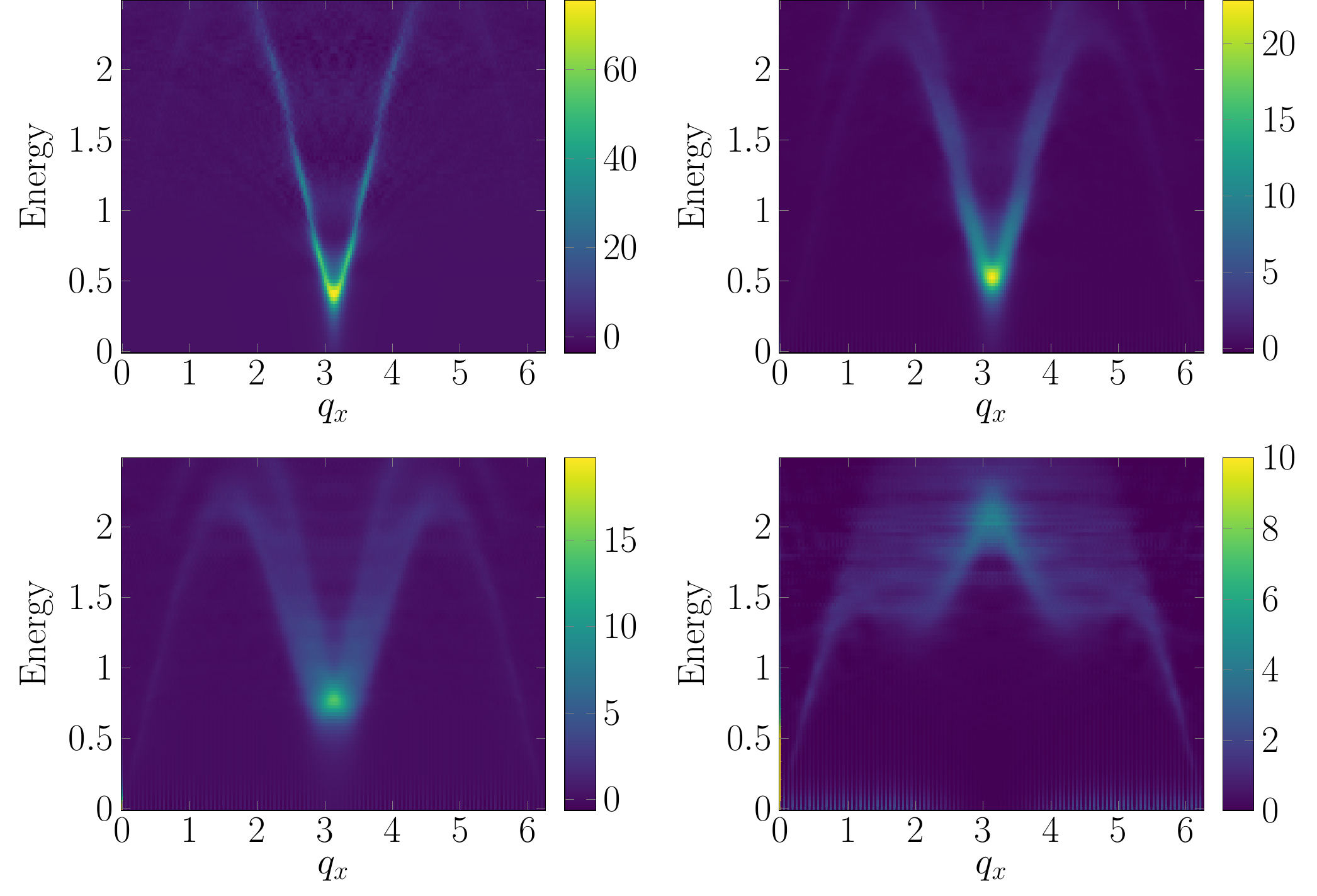}
\caption{\label{fig:haldaneSzszDynamics}First row: $S^{zz}(q,\omega)$ at field $h=0$
		and at field $h=0.5$ ($s_z \approx 0.05$).
		Last row:  $S^{zz}(q,\omega)$  at field $h=0.75$ ($s_z \approx 0.10$), and at field $h=2$ ($s_z \approx 0.40$).
		This is a 60-site DMRG run under OBC for the spin-1 Heisenberg model. }
\end{figure}
\begin{figure}
	\includegraphics[width=0.49\textwidth]{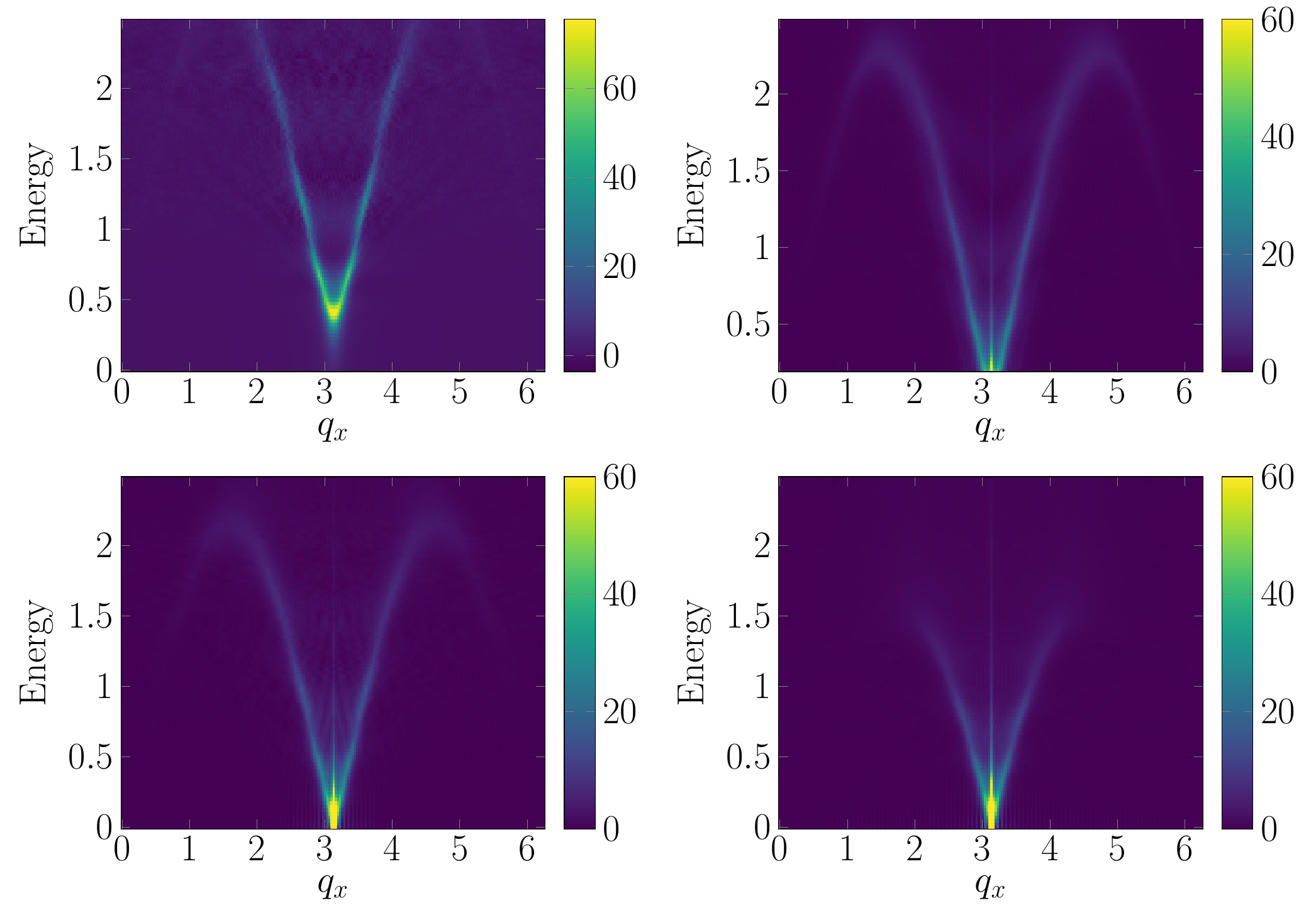}
\caption{\label{fig:haldaneSpsmDynamics}First row: $S^{+ -}(q,\omega)$ at field $h=0$
		and at field $h=0.5$ ($s_z \approx 0.05$).
		Last row:  $S^{+ -}(q,\omega)$  at field $h=0.75$ ($s_z \approx 0.10$), and at field $h=2$ ($s_z \approx 0.40$).
		This is a 60-site DMRG run under OBC for the spin-1 Heisenberg model.
}
\end{figure}

Figure \ref{fig:haldaneSzszDynamics} shows the $S^zS^z$ component of the dynamical structure factor calculated with the DMRG for a lattice of 60 sites with
open boundary conditions, with and without field $h$, as indicated.
Figure \ref{fig:haldaneSpsmDynamics} shows the $S^+S^-$ component of the dynamical structure factor calculated with the DMRG for a lattice of 60 sites with
open boundary conditions, with and without field $h$, as indicated. At $h < h_{c1}$ there is a non-zero Haldane gap as expected. The spin-1 Heisenberg model with zero field is exactly solvable via the bosonic perturbation on the Neel state. It gives a sharp dispersion of spin waves with a spectral gap $\Delta \sim 0.41$  \cite{Affleck_1989} which is consistent with results shown in Fig.\ref{fig:haldaneSpsmDynamics}(a).  

By comparing the spin-1 Heisenberg model results with those of the AKLT or the ULS,
we see that the quadratic term introduced by a not too large $\beta$ does not \emph{significantly}
affect the low energy spectrum of gapless intermediate phases. For example, the AKLT model with $\beta = 1/3$
retains most of the signature in its linearly-dispersed modes at $\pi$, and, as we show in the next subsection,
can be quantitatively captured by the single mode approximation with good accuracy.
Yet the effects of $\beta$ get stronger as it increases. As can be seen from the dynamical properties of the ULS model,
$\beta$ enhances the scattering of magnon excitations and results in the fractionalized signal seen
in the dynamical structure factor. This qualitatively agrees with the inaccuracy of NL$\sigma$M description of BLBQ for large
$\beta$ that it is incapable of capturing the correct gap in Haldane phase,
especially after crossing the AKLT point at $\beta = 1/3$ \cite{PhysRevB.102.014447}. However, as shown in the main text, such strong effect of the increment of $\beta$ on magnons, remarkably, is not as evidently reflected in its CFT description: although the change of $\beta$ strongly fractionalizes the magnon density of states, their effective CFT description with a central charge $c = 1$ remains the almost same.  

\section{Data Sources}
Data sources are distributed with supplemental. In file inputsPaper103.tar.xz we include input files to be delivered to DMRG++ for all figures in the main text and supplemental. In sourcesPaper103.tar.xz we include data and tex file for generating static and dynamical structure factors of ULS, AKLT and spin-1 Heisenberg model, and their SMA analysis and central charge. 

\section{How to reproduce the numerical results}
The \textsc{DMRG}++ computer program \cite{re:alvarez09} can be obtained with:

\begin{small}
	\begin{verbatim}
		git clone https://github.com/g1257/dmrgpp.git
	\end{verbatim}
\end{small}
and PsimagLite with:
\begin{small}
	\begin{verbatim}
		git clone https://github.com/g1257/PsimagLite.git
	\end{verbatim}
\end{small}
To compile:
\begin{small}
	\begin{verbatim}
		cd PsimagLite/lib; perl configure.pl; make
		cd ../../dmrgpp/src; perl configure.pl; make
	\end{verbatim}
\end{small}

The documentation can be found at 
\nolinkurl{https://g1257.github.io/dmrgPlusPlus/manual.html} or can be obtained
by doing \verb!cd dmrgpp/doc; make manual.pdf!.

The data needed for all the figures is in \verb!RawData.tar.gz!.

To reproduce figure 2, a ground state run needs to be made
with \verb!./dmrg -f akltGsL60Field0.ain! where all inputs are in  \verb!RawData.tar.gz!.
After the ground state run has finished,
the batches and inputs for all frequency runs can be generated with
\begin{widetext}
\begin{small}
	\begin{verbatim}
		export PSC=/path/to/dmrgpp/scripts
		perl -I${PSC} ${PSC}/manyOmegas.pl InputDollarizedAkltL60Field0.inp batchDollarized.pbs test
	\end{verbatim}
\end{small}
\end{widetext}
that can be launched by replacing \verb!test! with \verb!submit!.
When all frequency runs have finished, the post-processing to obtain the \verb!.pgfplots! 
files is (after setting \verb!PSC! as before)
\begin{widetext}
\begin{small}
    	\begin{verbatim}
		perl -I${PSC} ${PSC}/procOmegas.pl -f InputDollarizedAkltL60Field0.inp -p
	\end{verbatim}
\end{small}
\end{widetext}
At zero field, we can add the symmetry sector \verb!TargetSzPlusConst=! line, which should
equal to $Sz + L$, where $Sz$ is the symmetry sector to be targeted, and $L$ the number of sites.
We use $Sz$ plus the constant $L$ so that the number is always a non-negative integer.
At non-zero field, the local symmetry $Sz$ is also conserved, but we may not know what
value it should take, and therefore we do not provide the line \verb!TargetSzPlusConst=!
in the input file, so that the diagonalization includes all symmetry sectors.

\begin{widetext}
\section{Detailed Derivation of $\mathcal{C}(\beta)$}
Here we use the conventional BLBQ parameterized by $\beta$:
\begin{equation}
	H = \sum_{i,i+1} \left( \vec{S}_i\cdot \vec{S}_{i+1} \right) - \beta \left( \vec{S}_i\cdot \vec{S}_{i+1} \right)^2
\end{equation}
the commutator in SMA becomes
\begin{equation}
	\begin{split}
		\left[ S_{-q}^\alpha , [H_\theta, S_q^\alpha] \right]
		&= \sum_{inn'} \left[ S_{n'}^\alpha,\; \left[ \cos\theta \left( \vec{S}_i\cdot \vec{S}_{i+1} \right) + \sin\theta \left( \vec{S}_i\cdot \vec{S}_{i+1} \right)^2, \;S_n^\alpha \right] \right]\frac{e^{-iq(n-n')}}{L} \\
								      &= \sum_{inn'} \cos\theta \left( \left[S_{n'}^\alpha,  \vec{S}_i\cdot \vec{S}_{i+1}\right]S_n^\alpha - S_{n}^\alpha \left[ S_{n'}^\alpha, \vec{S}_{i}\cdot \vec{S}_{i+1} \right] \right)\frac{e^{-iq(n-n')}}{L} \\
								      &+ \sin\theta \left( \left[ S_{n'}^\alpha, \left( \vec{S}_i\cdot \vec{S}_{i+1} \right)^2 \right]S_n^\alpha - S_n^\alpha \left[ S_{n'}^\alpha, \left( \vec{S}_i\cdot \vec{S}_{i+1} \right)^2 \right] \right)\frac{e^{-iq(n-n')}}{L}
	\end{split}
\end{equation}
Next we evaluate the $\cos$(linear) and  $\sin$(quadratic) terms  seperately.

\noindent \textbf{Linear term}:
The linear cos part is
\begin{equation}
	\sum_{inn'} \left(\left[S_{n'}^z,  \vec{S}_i\cdot \vec{S}_{i+1}\right]S_n^z  - S_{n}^z \left[ S_{n'}^z, \vec{S}_{i}\cdot \vec{S}_{i+1} \right] \right)e^{-iq(n-n')}
\end{equation}
THe first term above gives
\begin{equation} \label{eq:2.4}
	\begin{split}
		-\sum_{inn'} &\left(-iS_i^x S_{n'}^y S_n^z\delta_{n',i+1} - iS_{n'}^y S_{i+1}^x S_n^z    \delta_{n',i} + iS_i^y S_{n'}^x S_n^z\delta_{n',i+1} + iS_{n'}^x S_{i+1}^y  S_n^z\delta_{n',i}\right)e^{-iq(n-n')}\\
			     &= \sum_{in} \left(iS_i^x S_{i+1}^y S_{n}^z e^{iq(i+1)} + iS_{i}^y S_{i+1}^x S_{n}^z e^{iqi}  - iS_{i}^y S_{i+1}^x S_{n}^ze^{iq(i+1)} - iS_i^x S_{i+1}^y  S_n^z e^{iqi} \right)e^{-iqn}
	\end{split}
\end{equation}
the second term is (including - sign)
\begin{equation} \label{eq:2.5}
	\begin{split}
		\sum_{inn'} &\left(-iS_n^z S_i^x S_{n'}^y\delta_{n',i+1} - iS_n^z S_{n'}^y S_{i+1}^x    \delta_{n',i} + iS_n^zS_i^y S_{n'}^x \delta_{n',i+1} + iS_n^zS_{n'}^x S_{i+1}^y \delta_{n',i}\right)e^{-iq(n-n')}\\
			     &= \sum_{in} \left(-iS_n^z S_i^x S_{i+1}^y e^{iq(i+1)} - iS_n^z S_{i}^y S_{i+1}^x e^{iqi}  + iS_n^z S_{i}^y S_{i+1}^x e^{iq(i+1)} + iS_n^z S_i^x S_{i+1}^y  e^{iqi} \right)e^{-iqn}
	\end{split}
\end{equation}
If $n \neq i$ and  $n \neq i+1$, then the commutation factor of cos term must vanish. So we only need to add up these two indices. This gives
\begin{equation}
	\begin{split}
		\sum_{i} &\left[ 2iS_i^x S_i^z S_{i+1}^y (\cos q -1) - 2iS_i^y S_i^z S_{i+1}^x (\cos q- 1) - 2iS_i^z S_i^x S_{i+1}^y (\cos q - 1) + 2iS_i^z S_i^y S_{i+1}^x (\cos q - 1)\right]\\
			 &= 2i(\cos q - 1) \sum_{i} \left( S_i^x S_i^z S_{i+1}^y - S_i^y S_i^z S_{i+1}^x - S_i^z S_i^x S_{i+1}^y + S_i^z S_i^y S_{i+1}^x \right) \\
			 &= 2i(\cos q - 1) \sum_{i} \left[S_i^x, S_i^z\right] S_{i+1}^y + \left[S_i^z, S_i^y\right] S_{i+1}^x \\
			 &= 2(\cos q - 1) \sum_{i} S_i^y S_{i+1}^y + S_i^x S_{i+1}^x 
	\end{split}
\end{equation}
This alone will be the SMA for the Haldane chain.

\vspace{+0.3cm}
\noindent\textbf{Quadratic term}
The sin term reads
\begin{equation}
	\sum_{inn'} \left( \left[ S_{n'}^z, \left( \vec{S}_i\cdot \vec{S}_{i+1} \right)^2 \right]S_n^z - S_n^z \left[ S_{ n'}^z, \left( \vec{S}_i\cdot \vec{S}_{i+1} \right)^2 \right] \right)e^{-iq(n-n')}
\end{equation}
In the same we we expand it  
\begin{equation}
	\begin{split}
		\sum_{in} &\left( \vec{S}_i\cdot \vec{S}_{i+1} \right) \left( iS_i^x S_{i+1}^y e^{iq(i+1)}+ iS_{i}^y S_{i+1}^x e^{iqi}- iS_i^y S_{i+1}^xe^{iq(i+1)} - iS_{i}^x S_{i+1}^ye^{iqi} \right)S_n^z e^{-iqn} \\
			  &+ \left( iS_i^x S_{i+1}^y e^{iq(i+1)} + iS_{i}^y S_{i+1}^x e^{iqi}- iS_i^y S_{i+1}^x e^{iq(i+1)} - iS_{i}^x S_{i+1}^y e^{iqi}\right) \left( \vec{S}_i\cdot \vec{S}_{i+1} \right) S_n^z e^{-iqn} \\
		 &- S_n^z \left( \vec{S}_i\cdot \vec{S}_{i+1} \right) \left( iS_i^x S_{i+1}^y e^{iq(i+1)} + iS_{i}^y S_{i+1}^x e^{iqi} - iS_i^y S_{i+1}^x e^{iq(i+1)} - iS_{i}^x S_{i+1}^y e^{iqi} \right) e^{-iqn} \\
			  &- S_n^z\left( iS_i^x S_{i+1}^y e^{iq(i+1)} + iS_{i}^y S_{i+1}^x e^{iqi} - iS_i^y S_{i+1}^x e^{iq(i+1)} - iS_{i}^x S_{i+1}^y e^{iqi} \right) \left( \vec{S}_i\cdot \vec{S}_{i+1} \right)  e^{-iqn} 
	\end{split}
\end{equation}
It's readily to see this equation evaluates to zero if $n \neq i, \: i+1$. In presence of inversion symmetry the summation becomes
\begin{equation}
	\begin{split}
		&\sum_{i} \left( \vec{S}_i\cdot \vec{S}_{i+1} \right) \Big( 2i(\cos q - 1) S_i^x S_i^z S_{i+1}^y  - 2i(\cos q - 1)S_i^y S_i^z S_{i+1}^x \Big)\\
		&= 2i (\cos q - 1) \sum_{i} \left( \vec{S}_i\cdot \vec{S}_{i+1} \right)\left( S_i^x S_i^z S_{i+1}^y - S_i^y S_i^z S_{i+1}^x \right) + \left( S_i^x S_{i+1}^y - S_i^y S_{i+1}^x \right) \left( \vec{S}_i\cdot \vec{S}_{i+1} \right)S_i^z - \\
		&- S_i^z \left( \vec{S}_i\cdot \vec{S}_{i+1} \right) \left( S_i^x S_{i+1}^y - S_i^y S_{i+1}^x \right) -  \left( S_i^z S_i^x S_{i+1}^y - S_i^z S_i^y S_{i+1}^x \right) \left( \vec{S}_i\cdot \vec{S}_{i+1} \right)
	\end{split}
\end{equation}
Here we make no attempt to simply the above equation because it is independent of momentum thus will not enter qualitatve behavior of dynamical structure. Hence we see the common factor $2(\cos\theta -1 )$ as is discussed in the main text, and is multiplied by the average correlators between nearest neighbors. The later is independent of momentum  $k$ so we can leave them off when studying the field dependence of the dispersion of a Hamiltonian. 
\end{widetext}

\bibliography{reference.bib}